\documentclass[11pt,a4paper]{article}
\pdfoutput=1
\usepackage{sint}
\usepackage{cite}
\usepackage{amsmath,amsfonts}
\usepackage{amssymb,bbm}
\usepackage{amsbsy,color}
\usepackage{mathrsfs}   
\usepackage{latexsym}
\usepackage{graphicx}
\usepackage{booktabs,dcolumn}
\usepackage{multirow}
\usepackage{multicol}
\usepackage[T1]{fontenc} 
\usepackage{macros_hs}
\definecolor{darkred}{rgb}{0.4,0.0,0.0}
\definecolor{darkgreen}{rgb}{0.0,0.4,0.0}
\definecolor{darkblue}{rgb}{0.0,0.0,0.4}
\usepackage[pdftex,a4paper,bookmarks,linktocpage,
    colorlinks,
    linkcolor = darkblue,
    urlcolor  = darkblue,
    citecolor = darkgreen
    ]{hyperref}
\begin{document}

\begin{titlepage}

\renewcommand{\thefootnote}{\fnsymbol{footnote}}

\begin{flushright}
\small{
IFIC/13-69 \\
DESY 13-231 \\
MS-TP-13-33 \\
}
\end{flushright}
\vspace{0.55cm}

\begin{center}
{\Large\bf
Matching of heavy-light flavour currents between HQET\\[0.5ex] 
at order $1/m$ and QCD:
I. Strategy and tree-level study
}
\end{center}
\vskip 0.35cm
\vbox{
\centerline{
\includegraphics[width=2.8cm]{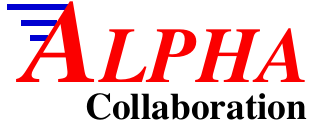}
}
}
\vskip 0.1cm
\begin{center}
{
Michele Della Morte$^{\scriptscriptstyle a}$,
Samantha Dooling$^{\scriptscriptstyle b}$\footnote[1]{%
Present address: Deutsches Elektronen-Synchrotron DESY, 
Notkestra{\ss}e 85, D-22607 Hamburg, Germany},
Jochen Heitger$^{\scriptscriptstyle b}$,
Dirk Hesse$^{\scriptscriptstyle c}$\\ 
and Hubert Simma$^{\scriptscriptstyle d}$
}
\vskip 0.5cm
{
\vskip 2.0ex
$^{\scriptstyle a}$
Instituto de F\'isica Corpuscular IFIC (CSIC),\\
c/ Catedr\'atico Jos\'e Beltr\'an 2, E-46980 Paterna, Spain
\vskip 2.0ex
$^{\scriptstyle b}$
Westf\"alische Wilhelms-Universit\"at M\"unster,
Institut f\"ur Theoretische Physik,
Wilhelm-Klemm-Stra{\ss}e 9, D-48149 M\"unster, Germany
\vskip 2.0ex
$^{\scriptstyle c}$
Universit\`a degli Studi di Parma,
Viale G.P. Usberti n.~7/A, I-43124 Parma, Italy 
\vskip 2.0ex
$^{\scriptstyle d}$
NIC, DESY,
Platanenallee 6, D-15738 Zeuthen, Germany
\vskip 2.0ex
}
\vskip 0.5cm
{\bf Abstract}
\vskip 0.1ex
\end{center}

We present a strategy how to match the full set of components of the 
heavy-light axial and vector currents in Heavy Quark Effective Theory (HQET), 
up to and including $1/m_{\rm h}$-corrections, to QCD.
While the ultimate goal is to apply these matching conditions 
non-perturbatively, in this study we first have implemented them at 
tree-level, in order to find good choices of the matching observables
with small $\Order{1/m_{\rm h}^2}$ contributions.
They can later be employed in the non-perturbative matching procedure
which is a crucial part of precision HQET computations of 
semileptonic decay form factors in lattice QCD.


\vskip 2.0ex

\vskip 2.0ex


\vfill
\eject

\end{titlepage}

\section{Introduction}
\label{sec:intro}
Experimental and theoretical studies of B-meson decays belong to the major
activities of the particle physics community within the realm of indirect 
searches for physics beyond the Standard Model.
In recent years, experimental advances have led to high-precision B-factory
experiments and LHCb, with the potential to observe signatures of 
New Physics by unveiling contributions to those weak decays owing to virtual 
effects of particles not present in the Standard Model.
These effects are known to be small and possibly quantifiable through the 
elements of the CKM matrix.
Its crucial ingredients are measured decay branching ratios on the 
experimental side and low-energy hadron matrix elements encoding the strong
interaction effects on the theory side. 
At the current high level of experimental precision, the r\^{o}le of equally 
accurate predictions from theory with controlled errors becomes increasingly 
important.
Lattice QCD provides an ab-initio non-perturbative approach for a reliable
and precise evaluation of these hadronic matrix elements. 

In the B-meson sector, a prominent example is the entry $|\Vub|$ of the CKM 
matrix, which can be estimated independently from the leptonic 
$\rmB \to \tau \nu$ and the exclusive semileptonic $\rmB \to \pi l \nu$ 
decay.
It has received much attention in the past, since indications for a tension 
at the level of about $3\sigma$ between these two determinations have been
reported (see, e.g., 
refs.~\cite{Lunghi:2010gv,Lenz:2012az,lat12:bphys,Rosner:2012np} 
and therein). 
Over the last few years, the ALPHA Collaboration has devised and implemented
a non-perturbative strategy for the computation of phenomenologically
relevant B-physics parameters in the framework of lattice Heavy Quark 
Effective Theory (HQET)~\cite{HQET:pap1,zastat:Nf2,HQET:param1m,HQET:msplit,HQET:fb1m,HQET:Nf2param1m}.
This strategy separates the b-quark mass scale from the other intrinsic 
scales in a lattice simulation by a systematic expansion in the inverse heavy
quark mass, $1/m_{\rm h}$, and is able to subtract the cumbersome ultraviolet 
power divergences of the effective theory (due to operator mixing under 
renormalization) such that the continuum limit exists.
While the low-energy constants of the effective theory (called HQET 
parameters from now on) entering the Lagrangian and the time component of the
heavy-light axial-vector current at order $1/m_{\rm h}$ have already been 
determined non-perturbatively~\cite{HQET:param1m,HQET:Nf2param1m} and applied 
to phenomenology~\cite{lat12:hqetNf2,lat13prlm:hqetNf2}, 
it is a natural and instructive next step to extend our HQET programme to also 
cover all components of the weak heavy-light axial and vector currents.
In particular, the QCD matrix elements of the vector current between
$\rmB$ ($\rmBs$) and $\pi$ ($\rmK$) states at finite momenta, entering the 
semileptonic processes $\rmB \to \pi l \nu$ and $\rmBs \to \rmK l \nu$, 
respectively, are parameterized by two form factors to be predicted by 
non-perturbative QCD and HQET.
Hence, determining the HQET parameters associated with the vector current 
including its $1/m_{\rm h}$-corrections will be an essential prerequisite for 
our current attempt at computing for the first time these semileptonic 
B-decay form factors by means of lattice HQET at 
$\Order{1/m_{\rm h}}$~\cite{lat12:fabio,ichep12:hqetNf2,HQET:Nf2semilep1m}.

The purpose of the present paper is twofold.
We first formulate a set of possible matching conditions for the HQET parameters
appearing in the Lagrangian and in all components of the heavy-light axial
($A_\mu$) and vector ($V_\mu$) currents including all $1/m_{\rm h}$-terms, 
which actually comprises terms of mass dimension five in the action and 
dimension four in the currents.
This results in a total of 19 HQET parameters and thus at least 19 matching 
equations required to determine them.
The second goal is to gain some insight into the intrinsic ambiguity of the
chosen matching conditions and to try to reduce that to a suitable level.
Let us expand on this point.
Each matching condition amounts to evaluate a certain  
observable in QCD and in HQET (including $1/m_{\rm h}$-terms) and to equate 
one with the other, in order to \emph{define} the parameters in HQET.
Ideally, one would like each observable to receive negligible contributions from 
$\Order{1/m_{\rm h}^2}$ terms and if possible
to be sensitive to a single
parameter only.
Otherwise, unnaturally large $\Order{1/m_{\rm h}^2}$ contributions 
may propagate into the physical quantities, which one ultimately wants
to compute in large-volume HQET after the matching step.
Indeed, as in previous applications of our general strategy,
the matching is here performed in finite volume with 
Schr\"odinger functional (SF) boundary conditions and linear extent 
$L$~\cite{SF:LNWW,SF:stefan1,SF:stefan2,zastat:pap1}.
Hence, our criterion for an optimal choice of matching observables is to
have them 
such that $(Lm_{\rm h})^{-n}$-corrections with $n\ge 2$ can be neglected.
These observables will then be considered as good candidates for the envisaged 
non-perturbative matching computation by numerical simulations.

We investigate in this paper the full system of 19 
matching equations at tree-level of perturbation theory, where the 
HQET parameters are known exactly and the  solution of the system 
can be studied as a function of $m_{\rm h}$ (and of further kinematical variables).
In this way we can estimate
the size of $1/m_{\rm h}^2$-contributions in the matching observables 
and in the HQET parameters.
The observables considered here are constructed from SF two- and three-point 
functions with appropriate kinematics.
For a one-loop study of the
renormalization factors of the static currents $A_k$ and $V_0$ based on
three-point functions, we refer to the companion paper~\cite{HQET:curr3pt1lp}.

The ALPHA Collaboration's B-physics programme is based on a non-perturbative 
matching of HQET to QCD in finite volume.
Let us briefly recall it for later convenience.
The HQET Lagrangian
\be
\lag{HQET}(x)=
\lag{stat}(x)+m_\mrm{bare}\psibarheavy(x)\,\psiheavy(x) 
-\omegakin\Okin(x)-\omegaspin\Ospin(x) \;,
\label{e:hqetact}
\ee
at leading order in $1/m_{\rm h}$ is just the (static) term
\be
\lag{stat}(x)=
\psibarheavy(x)\,D_0\,\psiheavy(x) \;
\label{e:statact}
\ee
(plus the mass term that only leads to a shift of the energy levels). 
At order $1/m_{\rm h}$, two additional interaction terms are included
\be
\Okin(x)=
\psibarheavy(x){\bf D}^2\psiheavy(x) \,,\quad
\Ospin(x)= 
\psibarheavy(x){\boldsymbol\sigma}\!\cdot\!{\bf B}\psiheavy(x) \,.
\label{e:okinspin}
\ee
They represent the kinetic energy from heavy quark's residual motion and the
chromomagnetic interaction with the gluon field. ${\bf D}^2$, ${\bf B}$, and $D_0$ are defined in ref.~\cite{HQET:param1m}.
Thus, the HQET Lagrangian has three parameters: $m_\mrm{bare}$, $\omegakin$, 
and $\omegaspin$. 
The predictivity of the effective theory is only granted, once these HQET
parameters have been fixed by a non-perturbative matching to 
QCD~\cite{HQET:pap1,Maiani:1992az}:
three (properly renormalized) QCD observables, $\Phi_i$ ($i=1,2,3$), 
evaluated in the continuum limit of \emph{finite-volume QCD}, 
are matched to their counterparts computed in HQET by imposing
\be
\Phi^\qcd_i(L,m_{\rm h},0)=
\Phi^\hqet_i(L,m_{\rm h},a) \,,
\label{e:matchingcond}
\ee
for any value of the lattice spacing $a$.
While the l.h.s. of this ``matching'' equation is defined in the continuum,
\be
      \Phi^\qcd_i(L,m_{\rm h},0) = \lim_{a\to 0}\Phi^\qcd_i(L,m_{\rm h},a)\;,
\ee
the quantities $\Phi^\hqet_i$ are understood to be expanded
up to a given order in $1/m_{\rm h}$ (NLO in our setup) and computed in HQET
at a finite lattice spacing.
By solving this system of (at this point three) matching equations, the resulting 
HQET parameters become functions of $m_{\rm h}$ and $a$ and can be pushed to 
lattice spacings for use in phenomenological applications with large-volume 
simulations by step-scaling methods.
For more details, the reader may consult, e.g., 
refs.~\cite{HQET:pap1,reviews:NPHQETrainer,HQET:param1m,HQET:Nf2param1m}, 
where this programme has been completed in the quenched approximation 
and for two flavours of $\Order{a}$ improved Wilson fermions.

For many relevant phenomenological applications in heavy quark physics, 
one also needs (correlation functions of) composite fields $\Oop^\qcd(x)$, 
which are local combinations of the fundamental fields.
The corresponding effective operators typically represent electroweak or 
other non-QCD interactions.
In HQET such operators are written as linear combinations
\be
\Oop^\hqet(m_{\rm h})= 
Z_O\left\{\Oop^\stat+\sum c_n\Oop_n\right\}=
\Oop^\qcd(m_{\rm h}) +\Order{1/m_{\rm h}^2} \,, 
\label{e:ohqet}
\ee
where the equality is meant for matrix elements of corresponding states in 
the fundamental (QCD) and effective (HQET) theories.
The r.h.s.~of the first equation above in general requires to include all 
operators which have a mass dimension one higher than $O^\qcd$ (or $\Oop^\stat$) 
and which transform in the same way as $O^\qcd$ under the common set of 
symmetries of QCD and HQET (but linearly independent with respect to the 
equations of motion of the effective theory).
The operators $\Oop_n$ are needed for the renormalization and $\Order{a}$ improvement 
of the effective theory, and in order to systematically include 
in HQET the $m_{\rm h}$-effects of QCD. These are encoded
into the non-trivial $m_{\rm h}$-dependence of the parameters $Z_O$ and $c_n$
in eq.~\ref{e:ohqet}. 
The remaining $m_{\rm h}$-dependence in 
$\Phi^\qcd_i(L,m_{\rm h},0)-\Phi^\hqet_i(L,m_{\rm h},a)$ is
useful to quantify higher-order (in $1/m_{\rm h}$) corrections.  
These will be studied here at tree-level for a particular class of matching 
conditions. 
In the following we assume all light quark masses to be set to zero,
and we shall only consider heavy-light currents,
$\Oop=\Jop\in\{A_0,A_k,V_0,V_k\}$, but the general strategy can
straightforwardly be applied to other operators, such as four-quark 
operators.
The HQET expansions of the currents at $\Order{1/m_{\rm h}}$ 
requires 16 additional HQET parameters. As it will be worked out 
in the next sections, they can be fixed on a similar footing as 
outlined for the 3 parameters of the Lagrangian above.
Note that the systematic expansion in $1/m_{\rm h}$ is part of the very 
definition of HQET and renders it order by order in $1/m_{\rm h}$ a
renormalizable field theory \cite{HQET:pap1}.

The plan of the paper is as follows.
In \sect{sec:s2} we explain the general structure of the system of matching 
equations for the parameters of the HQET Lagrangian together with heavy-light
currents and illustrate it for the case of the axial current as a
representative example.
\Sect{sec:s3} introduces the full set of matching observables as built from 
suitable finite-volume SF correlation functions and summarizes their generic
HQET expansions that enter the matching step;  
explicit expressions of the expansion coefficients are given
in appendix~\ref{sec:a1}. Our detailed discussion in the main text
is focused on the case of the axial current and the corresponding formulae 
for the vector current are collected in appendix~\ref{sec:a2}.
In \sect{sec:s4} our results for the HQET parameters from the tree-level
matching are discussed, and, based on that, we advocate our preferred 
choice of observables for the non-perturbative matching procedure in the 
future.   
\Sect{sec:concl} contains our conclusions.  
\vskip3em

\section{General form of the matching equations and strategy}
\label{sec:s2}

In order to determine all parameters, which occur at order $1/m_{\rm h}$ in
the HQET expansion (eq.~\ref{e:ohqet}) of a current $\Jop$,
the matching conditions in eq.~\ref{e:matchingcond} have to
be solved for a corresponding number of observables
$\Phi_i$ (say $i\in I_J$). 

These observables are constructed from suitable combinations 
of correlation functions, $C_J$, which typically have a single
insertion of $\Jop$. If the HQET expansion of $\Jop$ is written as
\be
    J^\hqet = Z^\hqet_J \left\{ \Jop^\stat + \sum_n c_{J_n} \Jop_n \right\}  + \Order{1/m_{\rm h}^2}\;,
    \label{e:jexpansion}
\ee
the HQET expansion of the correlation functions has the generic form
\bea
    C^\hqet_J =  Z^\hqet_J Z^\hqet_C e^{-m_\bare x_C} \left\{
                   C^\stat_J + \omegakin C^\kin_J + \omegaspin C^\spin_J + \sum_n c_{J_n} C^\nlo_{J_n} \right\}\;,
    \label{e:cfexpansion}
\eea
where all correlators on the r.h.s. are computed in the static
approximation.
The only place where the parameter $m_\bare$ appears is the factor $e^{-m_\bare x_C}$, 
with $x_C$ related to the time distances of the heavy (static) quark fields 
entering in $C^\hqet_J$.
Aside from 
$Z^\hqet_J$, all other (wave-function) renormalization factors contributing
to $C^\hqet_J$ are collected in $Z^\hqet_C$.

In the correlation functions on the r.h.s. of eq.~\ref{e:cfexpansion},
the leading-order term $C^\stat_J$ has just one insertion of $\Jop^\stat$
(instead of $\Jop$), while $C^{\kin/\spin}_J$ differ from $C^\stat_J$ by an 
extra insertion (summed over the entire space-time volume) of the $1/m_{\rm h}$-terms 
$\Okin$ or $\Ospin$ from the Lagrangian. 
The other next-to-leading contributions $C^\nlo_{J_n}$ have an insertion of 
one of the higher-dimensional operators $\Jop_n$
from the expansion in eq.~\ref{e:jexpansion}. 

The observables $\Phi_i$ are then defined as suitable combinations of such correlation 
functions (e.g., logarithms of ratios, see section~\ref{sec:s3} for explicit definitions).
Thus, the renormalization factor $Z^\hqet_C$ cancels, while a possibly remaining 
factor $Z^\hqet_J$ needs to be expanded in $1/m_{\rm h}$ in order to consistently
keep only terms up to order $1/m_{\rm h}$ in eq.~\ref{e:cfexpansion}. Moreover, only 
one of the observables, say $\Phi_1$, is left with an explicit dependence on $m_\bare$.

By combining all HQET parameters into a vector 
\be
   \omega = (m_\bare,\ \omegakin,\ \omegaspin,\ c_{J_1},\ \ldots,\ \ln Z^{\rm HQET}_{J},\ \ldots )^T \ ,
\ee
the HQET expansion of the observables can be written in the compact form
\bea
 \Phi_i^\hqet(L,M,a) = \eta_i(L,a) + \varphi_i^j(L,a)\,\omega_j(M,a) + \Order{1/m_{\rm h}^2}\,, 
 \label{e:phiexp}
\eea
where $M$ is the Renormalization Group Invariant (RGI) heavy quark mass and
the vector $\eta$ accounts for the contribution of the static terms $C_J^{\rm stat}$ in the
correlators involved. 
As mentioned in the introduction, because the matching is performed at a finite value of 
the renormalized quark mass of QCD, 
the parameters get a non-trivial $M$-dependence even when working only in the static 
approximation of HQET. 
Note that in general and non-perturbatively, only the combination of HQET 
quantities, which enters on the r.h.s. of eq.~\ref{e:phiexp}, is expected to 
have a continuum limit. The tree-level approximation is exceptional in this respect as 
each individual term on the r.h.s. has a well defined continuum limit in that case.

The matrix $\varphi_i^j$ of eq.~\ref{e:phiexp} reflects the structure of the matching
equations \ref{e:matchingcond}.
To illustrate its general structure, we group the parameters into
blocks $(m_\mrm{bare})$, $(\omegakin,\,\omegaspin)$,
 $(c_{J_{i}},\,Z_J )$, $(c_{J^\prime_{i}},\,Z_{J^\prime} )$,
$(c_{J^{\prime\prime}_{i}},\,Z_{J^{\prime\prime}} )$, \ldots,
and assume that $J$ is the current which is used in $\Phi_1$.
A suitable choice of the other observables then leads to the following natural form of
the matrix $\varphi_i^j$:
\bi
\item In the {\bf first column} all entries, except for $\varphi_1^1$, vanish.
\item The {\bf first row} has non-vanishing $\varphi_1^2$ and $\varphi_1^3$ corresponding 
      to contributions from $\omegakin$ and $\omegaspin$ to $\Phi_1$.
      In addition, there may be non-zero $\varphi_1^j$ with $j$ from a single
      block, which corresponds to the current used in $\Phi_1$.
      In our case, this is due to the  $A_{0,1}$-term, which 
      enters in our matching condition for $m_\mrm{bare}$ (see later). However,  
      it is easy to show that the corresponding
      $\varphi_1^4$ would vanish in the large-$L$ limit.
\item The rest has a simple {\bf block structure}, with non-zero blocks
      only in the second block-column (corresponding to contributions from
      $\omegakin$ and $\omegaspin$) and in the blocks on the diagonal 
      (corresponding to the mixings in the last term of 
      eq.~\ref{e:cfexpansion}).
\ei
Thus, 
we schematically
have the following block structure:
\begin{displaymath}
  \varphi = \left(
     \begin{array}{c|c|c|c|c}
         \varphi_1^1 & \ast & \ast  &   0   &    0    \\ \hline
         0            & \ast &   0   &   0   &   0    \\ \hline
         0            & \ast & \ast  &   0   &   0    \\ \hline
         0            & \ast &   0   & \ast  &   0    \\ \hline
         0            & \ast &   0   &   0   & \ast   \\
     \end{array}
  \right)\;.
\end{displaymath}
Each time  an additional (effective) operator $\Jop^\prime$ is included, 
a new set of observables $\Phi_i$ with $i\in I_{J^\prime}$ can be added 
such that $\varphi_i^j$, and hence the matching equations, have the
above block structure. The  system can therefore always be solved simply 
by block-wise backward substitution in order to determine the  HQET 
parameters.


\subsection{Heavy-light axial current in HQET}

As an example of this general structure of the matching equations,
we recall the explicit form of the (renormalized) heavy-light 
axial current in HQET. The time component is
\bea
  \label{e:ahqet}
  A^\hqet_0(x) = Z_{{\rm A}_0}^\hqet\left[\,A^\stat_0(x)+\sum_{i=1}^2\ceff{A}{0}{i} A_{0,i}(x)\,\right] \,,
\eea
with the leading-order (static) term
\bea
 A^\stat_\mu(x)  & = & \psibarlight(x)\gamma_\mu\gamma_5\psiheavy(x)
\eea
and two additional dimension-four contributions
\bea
 A_{0,1}(x) &=& \psibarlight(x)\frac{1}{2}
            \gamma_5\gamma_i(\nabsym{i}-\lnabsym{i})\psiheavy(x)\,,
\\
 A_{0,2}(x) &=& \psibarlight(x)\frac{1}{2}
            \gamma_5\gamma_i(\nabsym{i}+\lnabsym{i})\psiheavy(x)\,,
\eea
where all derivatives are symmetric,
\be
  \drvsym{i}  = \frac12(\drv{i}+\drvstar{i})\,,\quad
  \lnabsym{i} = \frac12(\lnab{i}+\lnabstar{i})\,,\quad
  \nabsym{i}  = \frac12(\nab{i}+\nabstar{i})\,.
\ee

A comment on the definition of the covariant derivatives is in order here.
On a finite lattice, the action of the covariant derivative on a fermion field is given by
\begin{eqnarray}
\nab{\mu}\psi(x)&=&{{1}\over{a}} \left[ \lambda_\mu U(x,\mu)\psi(x+a\hat{\mu})-\psi(x)\right]\;, \\
\nabstar{\mu}\psi(x)&=& {{1}\over{a}}\left[ \psi(x) -\lambda_\mu^{-1}U(x-a\hat{\mu},\mu)^{-1}\psi(x-a\hat{\mu}\right]\;,
\end{eqnarray}
where $\hat{\mu}$ is a unit vector in direction $\mu$. The left action is defined as
\begin{eqnarray}
\overline{\psi}(x) \lnab{\mu}&=&{{1}\over{a}} \left[\, \overline{\psi}(x+a\hat{\mu})U(x,\mu)^{-1}\lambda_\mu^{-1}- \overline{\psi}(x)\right]\;, \\
\overline{\psi}(x) \lnabstar{\mu}&=&{{1}\over{a}}\left[\,\overline{\psi}(x) -\overline{\psi}(x-a\hat{\mu})U(x-a\hat{\mu},\mu)\lambda_\mu\right]\;.
\end{eqnarray}
We want to emphasize here the appearance of the phase factor
\be
\lambda_\mu=e^{ia\theta_\mu/L}\;, \quad \theta_0=0\;, \quad -\pi < \theta_k \leq \pi\;,
\ee
which  plays an important r\^ole in the following, as it will be used to vary the kinematics
in the correlation functions. Up to an Abelian gauge transformation,
it is equivalent to imposing the generalized periodic boundary conditions 
\be
\psi(x+L\hat{k})=e^{i\theta_k}\psi(x)\;, \qquad \overline{\psi}(x+L\hat{k})=\overline{\psi}(x)e^{-i\theta_k}\;.
\label{e:theta}
\ee
In our setup, this phase, sometimes referred to as {\em twisting}~\cite{Sachrajda:2004mi},
 can be employed to inject a momentum $|\Th -\Tq|/L$
in the correlation functions we describe in the next section.
It is clear that for $\Th=\Tq$ composite fields as $A_{0,2}(x)$ above can be associated with total derivative operators. 
In such a case their contribution vanishes once inserted
in correlation functions, unless a non-zero total momentum component (integer multiple of $2\pi/L$) is
explicitly considered by introducing spatial integrations with the appropriate Fourier factors
(which we avoid to do, in order to fully exploit volume averaging in view of 
numerical applications).

Concerning the spatial components of the axial-vector current, their
HQET expansion (see ref.~\cite{reviews:NPHQETrainer}) can be written as
\be
  A^\hqet_k(x) = Z_{\vec{\rm A}}^\hqet\left[\,A_k^\stat(x)+ \sum_{i=1}^4 c_{{\rm A}_{k,i}} A_{k,i}(x)\,\right]\,, 
  \label{e:vhqet}
\ee
with the following four extra terms:
\bea
\nonumber
 A_{k,1}(x) & = & \psibarlight(x){1\over2}
            (\nabsym{i}-\lnabsym{i}) \gamma_i\gamma_5\gamma_k  \psiheavy(x)\,,
\\
\nonumber
 A_{k,2}(x) & = & \psibarlight(x){1\over2}
            (\nabsym{k}-\lnabsym{k})\gamma_5\psiheavy(x)\,,
\\
\nonumber
 A_{k,3}(x) & = & \psibarlight(x){1\over2}
            (\nabsym{i}+\lnabsym{i}) \gamma_i\gamma_5\gamma_k  \psiheavy(x)\,,
\\
\nonumber
A_{k,4}(x) & = & \psibarlight(x){1\over2}
            (\nabsym{k}+\lnabsym{k})\gamma_5\psiheavy(x)\,.
\eea
The vector current components are just obtained by dropping $\gamma_5$ in these
expressions and by changing $\ceff{A}{\mu}{i}\to \ceff{V}{\mu}{i}$.
The classical values of the coefficients are
\be
\ceff{A}{0}{1}=\ceff{A}{0}{2}=-\ceff{A}{k}{1}=-\ceff{A}{k}{3}
=-{1 \over 2m_{\rm h}} \;, 
\quad {\rm whilst} \quad 
\ceff{A}{k}{2}=\ceff{A}{k}{4}={1 \over m_{\rm h}} \;,
\ee
and analogous for the vector current, apart from
$\ceff{V}{0}{1}$, $\ceff{V}{0}{2}$ and  $\ceff{V}{k}{2}$, $\ceff{V}{k}{4}$ which differ by a 
sign.
As a remark, the heavy-light pseudoscalar and scalar densities $P$ and $S$ 
are given exactly by the same expressions as $V_0$ and $A_0$ 
(due to the $\gamma_0 \psi_{\rm h}=\psi_{\rm h}$ property) with the changes 
$\ceff{A}{0}{i} \to c_{{\rm P}_i}$, $\ceff{V}{0}{i} \to c_{{\rm S}_i}$.
Such coefficients can be determined essentially through the same matching 
conditions
we are going to implement for $A_0$ and $V_0$, using instead correlation
functions involving $P$ and $S$ on the QCD side.

We note in passing that the mixing of $A_k^\stat$ with $A_{k,2}$ and $A_{k,4}$ 
is due to the breaking of spin-symmetry by $\Ospin$.
Similarly, the mixing of the static currents with the (combinations of) 
operators, like $A_{0,1}+A_{0,2}$, $A_{k,1}+A_{k,3}$, and $A_{k,2}+A_{k,4}$, 
where the derivative only acts on the heavy quark, are due to the 
breaking of local heavy-flavour conservation by $\Okin$. 

In refs.~\cite{HQET:param1m,HQET:Nf2param1m} only the time component 
of the axial current has been considered,
and $A_{0,2}$ has not been included because it does not contribute to 
correlation functions at (total) zero momentum, such as those typically 
used to compute decay constants.
The corresponding 5 HQET parameters can then be determined from a restricted
system of matching equations which has the form (see ref.~\cite{HQET:param1m}
for a precise definition of the quantities in the matrix below)
\begin{displaymath}
   \def\dA{{\delta {\rm A}}}
   \def\cA{{c^{(1)}_{\rm A}}}
   \left(\begin{array}{c}
     L \Gamma^{\rm P} \\[1mm]
     \hline \\[-2mm]
     R_1    \\[1mm]
     \frac{3}{4} \ln\left(\frac{f_1}{k_1}\right) \\[2mm]
     \hline \\[-2mm]
     R_{\rm A}    \\[2mm]
     \ln\left(\frac{-f_{\rm A}}{\sqrt{f_1}}\right) 
   \end{array}\right)
   = 
   \left(\begin{array}{c}
     L \Gamma^\stat \\[2mm]
     \hline   \\[-2mm]
     R_1^\stat \\[2mm]
     0        \\[2mm]
     \hline   \\[-2mm]
     R_{\rm A}^\stat \\[2mm]
     \zeta_{\rm A}  \\[2mm]
   \end{array}\right)
   + 
   \left(\begin{array}{c|cc|ccc}
     L & L\Gamma^\kin & L\Gamma^\spin    &  L\Gamma_\dA & 0 \\[2mm]
    \hline &&&&& \\[-2mm]                                         
     0 & R_1^\kin     &       0          &       0      & 0 \\[2mm]
     0 &    0             &  \rho_1^\spin    &       0      & 0 \\[2mm]
    \hline &&&&& \\[-2mm]                                         
     0 & R_{\rm A}^\kin     &  R_{\rm A}^\spin       &   R_\dA      & 0 \\[2mm]
     0 & \psi^\kin    & \psi^\spin       &   \rho_\dA   & 1 \\
   \end{array}\right)
   \cdot
   \left(\begin{array}{c}
      m_{\rm bare}        \\[2mm]
    \hline\\[-2mm]
      \omegakin          \\[2mm]
      \omegaspin         \\[2mm]
    \hline\\[-2mm]
      a \cA              \\[2mm]
      \ln{Z_{\rm A}^{\rm HQET}} \\[2mm]
   \end{array}\right)\;.
\end{displaymath}

\section{Matching of heavy-light currents at $\Order{1/m_{\rm h}}$}
\label{sec:s3}

We now consider the full system of 19 matching equations for HQET including the 
currents $J = A_0$, $A_k$, $V_0$, $V_k$. 
We arrange the parameters in a vector $\omega \equiv \omega_i$ as follows:
\begin{center}
\renewcommand{\arraystretch}{1.25}
\begin{tabular}{@{\extracolsep{0.2cm}}lll}
\toprule
      $i$ & $\omega_i$ & origin\\ 
      \midrule \hline
      1, 2, 3 & $m_\bare,\ \omegakin,\ \omegakin$                           & $\lag{HQET}$  
      \\[2mm] 
      4, \ldots, 6 & $\ceff{A}{0}{1},\ \ceff{A}{0}{2},\ \ln Z_{A_0}^\hqet$  & $A^\hqet_0$
      \\[2mm]
      7, \ldots, 11 & $\ceff{A}{k}{1},\ \ceff{A}{k}{2},\ \ceff{A}{k}{3},\ \ceff{A}{k}{4},\ \ln Z_{\vec{\rm A}}^\hqet$ & $A^\hqet_k$   
      \\[2mm]
      12 \ldots, 14 & $\ceff{V}{0}{1},\ \ceff{V}{0}{2},\ \ln Z_{V_0}^\hqet$ & $V^\hqet_0$
      \\[2mm]
      15, \ldots, 19 & $\ceff{V}{k}{1},\ \ceff{V}{k}{2},\ \ceff{V}{k}{3},\ \ceff{V}{k}{4},\ \ln Z_{\vec{\rm V}}^\hqet$ & $V^\hqet_k$  \\ 
\bottomrule
\end{tabular}
\end{center}
where in the last column we have indicated for each parameter, whether it enters the HQET Lagrangian
or the expansion of a current component. We limit our more detailed discussion to the 
parameters in $\lag{HQET}$, $A_0$ and $A_k$ (i.e., $\omega_i$ with $1\leq i \leq 11$).
The matching equations for $V_0$ and $V_k$ are simply obtained by generalizing those
for $A_0$ and $A_k$, and further details can be found in appendix~\ref{sec:a2}.

\subsection{Definition of the correlation functions in QCD}
As in previous work \cite{HQET:pap1,HQET:param1m,HQET:Nf2param1m}, we define 
the matching observables in the SF with homogeneous boundary conditions at $x_0=0$ and $x_0=T$
\cite{SF:LNWW,SF:stefan1,SF:stefan2}.
Correlation functions can be formed from composite fields in the bulk, $0<x_0<T$, and boundary quark fields.
We obviously think of $m_{\rm h}$ being around the mass of the b-quark and
use the label b to refer to heavy {\em relativistic} quarks.
The subscript I of the currents indicates that they are ${\mathcal{O}}(a)$ improved, 
as defined in ref.~\cite{impr:pap1,impr:pap5} for Wilson quarks.
We define in QCD on the lattice:
\begin{itemize}
\item Boundary-to-boundary correlators
   \bea
   \label{e:cffirst}
   \Fone
   (\T_\ell,\thb) & = &  -{a^{12} \over 2L^6}\sum_{\vecu,\vecv,\vecy,\vecz}
     \left\langle
     \zetabarprime_\light(\vecu)\gamma_5\zetaprime_{\rm b}(\vecv)\,
     \zetabar_{\rm b}(\vecy)\gamma_5\zeta_\light(\vecz)
     \right\rangle\,, \\
   \Kone
   (\T_\ell,\thb) & = &
     -{a^{12} \over 6L^6}\sum_{k}\sum_{\vecu,\vecv,\vecy,\vecz}
     \left\langle
     \zetabarprime_\light(\vecu)\gamma_k\zetaprime_{\rm b}(\vecv)\,
     \zetabar_{\rm b}(\vecy)\gamma_k\zeta_\light(\vecz)
     \right\rangle\,, \\
   K^{\light\light}_1(\vec{\theta}_{\ell},\vec{\theta}_{\ell'}) & = &
     -{a^{12} \over 6L^6}\sum_{k}\sum_{\vecu,\vecv,\vecy,\vecz}
     \left\langle
     \zetabarprime_\light(\vecu)\gamma_k\zetaprime_{\ell'}(\vecv)\,
     \zetabar_{\ell'}(\vecy)\gamma_k\zeta_\light(\vecz)
     \right\rangle\,. \\
   \j1a1(x_0,\T_{\ell},\T_{\ell'},\T_{\rm b}) & = &
    -{{a^{15}}\over{2L^6}}\sum_{\bf u,v,y,z,x} \langle \zetabar'_{\ell'}({\bf u}) \gamma_1 \zeta'_{\ell} ({\bf v})
    (A_{\rm I})_1(x) \zetabar_{\rm b} ({\bf z})\gamma_5 \zeta_{\ell'} ({\bf y})\rangle \;.
   \eea

\item Bulk-to-boundary correlators
\bea
\fa
(x_0,\T_\ell,\thb) & = & -{a^6 \over 2}\sum_{\vecy,\vecz}\,
  \left\langle
  (\aimpr)_0(x)\,\zetabar_{\rm b}(\vecy)\gamma_5\zeta_\light(\vecz)
  \right\rangle  \,, \label{e_fa}
\\
\fak
(x_0,\T_\ell,\thb) & = & i{a^6 \over 6}\sum_{k}\sum_{\vecy,\vecz}\,
  \left\langle
  (\aimpr)_k(x)\,\zetabar_{\rm b}(\vecy)\gamma_5\zeta_\light(\vecz)
  \right\rangle  \,, \label{e_fak}
\\
\ka21
(x_0,\T_\ell,\thb) & = & i{a^6 \over 2}\sum_{\vecy,\vecz}\,
  \left\langle
  (\aimpr)_2(x)\,\zetabar_{\rm b}(\vecy)\gamma_1\zeta_\light(\vecz)
  \right\rangle  \,, \label{e_fav21} 
  \label{e:cflast}
\eea
where for the last two correlators the $\T$ angles have to be chosen different from zero, 
at least in the $z$-direction, as otherwise they would vanish due to either rotation invariance or parity.
\ei

The inclusion of the static quarks has been discussed in ref.~\cite{zastat:pap1}, 
to which we refer for the definition of the boundary quark fields $\zeta_{\rm h}$ 
and $\zetabar_{\rm h}$. Taking $\fa$ as an example, we define the correlation functions 
that enter in the $1/m_{\rm h}$-expansion of HQET:
\bea
\fa^\stat(x_0,\Tq) & = & -{a^6 \over 2}\sum_{\vecy,\vecz}\,
  \left\langle
  A_0^\stat(x)\,\zetabar_\heavy(\vecy)\gamma_5\zeta_\light(\vecz)
  \right\rangle\;,
\\
\fa^\spin(x_0,\Tq) & = & -{a^{10} \over 2}\sum_{\vecy,\vecz,w}\,
  \left\langle
  \Ospin(w) A_0^\stat(x)\,\zetabar_\heavy(\vecy)\gamma_5\zeta_\light(\vecz)
  \right\rangle\;,
\eea
\bea
\fa^\kin(x_0,\Tq,\Th) & = & -{a^{10} \over 2}\sum_{\vecy,\vecz,w}\,
  \left\langle
  \Okin(w) A_0^\stat(x)\,\zetabar_\heavy(\vecy)\gamma_5\zeta_\light(\vecz)
  \right\rangle\;,
\\
f_{{\rm A}_{0,i}}(x_0,\Tq,\Th) & = &
-{a^6 \over 2}\sum_{\vecy,\vecz}\,
  \left\langle
  A_{0,i}(x)\,\zetabar_\heavy(\vecy)\gamma_5\zeta_\light(\vecz)
  \right\rangle\;.
\eea
We will use an analogous notation for all the HQET correlators appearing
in the expansion of the QCD correlators defined in eqs.~\ref{e:cffirst} -- \ref{e:cflast} 
and in appendix~\ref{sec:a2}.
For the renormalized correlator $\left[ \fa \right]_{\rm R}$, the HQET expansion 
eq.~\ref{e:cfexpansion} including order $1/m_{\rm h}$-terms reads
\be
  \left[\fa\right]^\hqet_{\rm R}  =  
    Z_{{\rm A}_0}^\hqet Z_{\zeta_{\rm h}} Z_\zeta e^{-m_{\rm bare}x_0} 
    \left\{ f_{{\rm A}_0}^\stat + \omega_{\rm kin} \fa^\kin + \omega_{\rm spin} \fa^\spin 
    + \sum_i c_{{\rm A}_{0,i}} f_{{\rm A}_{0,i}} \right\} \;.
\ee
Notice that there are no $1/m_{\rm h}$-terms arising from the HQET expansion of the boundary fields.
These indeed vanish or can be absorbed into the field-normalization,
either because of the equations of motion or by
imposing the SF boundary condition 
\be
P_-\zeta_{\rm b}=0\;, \qquad P_-=\frac{1}{2} \left( 1-\gamma_0\right)\;,
\ee
to all orders in $1/m_{\rm h}$.
\subsection{Choice of the observables}
\label{subsec:obs}
We now show that the correlators defined above are sufficient to form
a set of rather simple observables, which can be employed to fix the HQET 
Lagrangian and the axial current components at $\Order{1/m_{\rm h}}$.
We suppress the arguments $L,\; M$ and $a$ with respect to eq.~\ref{e:matchingcond} 
and rather emphasize the dependence on the fermionic
periodicity angles entering the different correlators.
For the determination of the parameters $\omega_i$, $1\leq i \leq 11$, 
we define the following 11 observables:
\def\phiseparator{8mm}
\begin{displaymath}
  \begin{array}{lcll}
  \Phi_1^\qcd(\T_1)  & \equiv & - L \cdot \tilde{\partial}_0 
        \ln \left(-\fa(x_0,\T_1,\T_1)\right)\,, 
  & (T=L,\ x_0=T/2)\,,
  \\[\phiseparator]
  \Phi_2^\qcd\Atwo & \equiv & \displaystyle 
        \frac{1}{4} \ln\left(\frac{F_1(\T_1,\T_1)}{F_1(\T_2,\T_2)}\right)
      + \frac{3}{4} \ln\left(\frac{K_1(\T_1,\T_1)}{K_1(\T_2,\T_2)}\right)\,,
  & (T=L/2)\,,
  \\[\phiseparator]
  \Phi_3^\qcd(\T_1)  & \equiv & \displaystyle 
        \frac{3}{4} \ln\left(\frac{F_1(\T_1,\T_1)}{K_1(\T_1,\T_1)}\right)\,,
  & (T=L/2)\,,
  \\[\phiseparator]
  \Phi_4^\qcd\Atwo & \equiv & \displaystyle
      \ln\left(\frac{f_{\rm A_0}(x_0,\T_1,\T_1)}{f_{\rm A_0}(x_0,\T_2,\T_2)}\right)\,, 
  & (T=L,\ x_0=T/2)\,,
  \\[\phiseparator]
  \Phi_5^\qcd\Athree & \equiv & \displaystyle
      \ln\left(\frac{f_{\rm A_0}(x_0,\T_1,\T_2)}{f_{\rm A_0}(x_0,\T_1,\T_3)}\right) \,,
  & (T=L,\ x_0=T/2)\,,
  \\[\phiseparator]
  \Phi_6^\qcd(\T_1) & \equiv & \displaystyle
      \ln\left(\frac{-f_{\rm A_0}(x_0,\T_1,\T_1)}{\sqrt{F_1(\T_1,\T_1)}}\right)\,, 
  & (T=L,\ x_0=T/2)\,,
\end{array}
\end{displaymath}
\begin{displaymath}
  \begin{array}{lcll}
  \Phi_7^\qcd\Atwo & \equiv & \displaystyle 
      \ln\left(\frac{f_{\vec{\rm A}}(x_0,\T_1,\T_1)}{f_{\vec{\rm A}}(x_0,\T_2,\T_2)}\right)\,,
  & (T=L,\ x_0=T/2)\,,
  \\[\phiseparator]
  \Phi_8^\qcd\Atwo & \equiv & \displaystyle
      \ln\left(\frac{k^1_{\rm A_2}(x_0,\T_1,\T_1)}{k^1_{\rm A_2}(x_0,\T_2,\T_2)}\right)\,,
  & (T=L,\ x_0=T/2)\,,
  \\[\phiseparator]
  \Phi_9^\qcd\Athree & \equiv & \displaystyle 
      \ln\left(\frac{f_{\vec{\rm A}}(x_0,\T_1,\T_2)}{f_{\vec{\rm A}}(x_0,\T_1,\T_3)}\right)\,,
  & (T=L,\ x_0=T/2)\,,
  \\[\phiseparator]
  \Phi_{10}^\qcd\Athree & \equiv & \displaystyle 
      \ln\left(\frac{k^1_{\rm A_2}(x_0,\T_1,\T_2)}{k^1_{\rm A_2}(x_0,\T_1,\T_3)}\right)\,,
  & (T=L,\ x_0=T/2)\,,
  \\[\phiseparator]
  \Phi_{11}^\qcd(\T_1) & \equiv & \displaystyle 
      \ln\left(\frac{J_{\rm A_1}^1(x_0,\T_1,\T_1,\T_1)}{\sqrt{F_1(\T_1,\T_1) \times K^{\light\light}_1(\T_1,\T_1)}} \right)\,,
  & (T=L,\ x_0=T/2)\,.
\end{array}
\end{displaymath}
We prefer to determine the normalization factor of the spatial components of the axial current
through $\Phi_{11}^\qcd$ which uses a boundary-to-boundary correlation function (corresponding 
to a three-point function in large volume) as proposed in \cite{HQET:curr3pt1lp}.
Alternatively, we could also use
\be
\Phi_{11}^{\prime\,\qcd}(\T_1)  \equiv  \displaystyle 
      \ln\left(\frac{f_{\vec{\rm A}}(x_0,\T_1,\T_1)}{\sqrt{F_1(\T_1,\T_1)}}\right)\,,
  \qquad\qquad (T=L,\ x_0=T/2)\,, 
\label{e:Phi11old}
\ee
which is defined entirely in terms of boundary-to-bulk correlators (two-point functions).
Our results at tree-level indicate that $\Phi_{11}^\qcd$ yields smaller 
higher-order corrections in $1/m_{\rm h}$ (see section~\ref{sec:s4}).
This holds true already at the static order where spin-symmetry is exact. For this reason we will preferably
adopt a three-point function also for the normalization of the temporal component of the vector current
(see the definition of $\Phi^{\rm QCD}_{14}$ in appendix~\ref{sec:a2} and ref.~\cite{HQET:curr3pt1lp} for a 
perturbative study at one-loop). The list of the additional
correlators and the observables $\Phi_{12}^\qcd,\dots,\Phi_{19}^\qcd$ needed 
for the matching of the temporal and spatial components of the heavy-light 
vector current are contained in appendix~\ref{sec:a2}.  

The matching equations just state that the above observables
are equal to the corresponding HQET counterparts. 
The explicit form of the HQET expansions according to eq.~\ref{e:phiexp} 
will be discussed in the next subsection. 
Here, we like to point out that the freedom to select a specific kinematics 
through the $\T$ angles of eq.~\ref{e:theta} is crucial in our choice of 
the matching observables. The same combination of correlation function, but with 
different kinematics, can provide sensitivity to different HQET parameters.
For example, $\Phi_4^\qcd$ and $\Phi_5^\qcd$ differ only in their kinematics.
In this way the HQET expansion of $\Phi_4^\hqet$ does not receive contributions 
from correlators with insertions of the total derivative operator $A_{0,2}$, while 
$\Phi_5^\qcd$ does.
In other (slightly abusing) words, $\Phi_4^\qcd$ is sensitive to 
$c_{{\rm A}_{0,1}}$, while $\Phi_5^\qcd$ is sensitive to both $c_{{\rm A}_{0,1}}$ 
and $c_{{\rm A}_{0,2}}$.
Similarly, $\Phi_7^\qcd$ and $\Phi_8^\qcd$ are sensitive to
$c_{{\rm A}_{k,1}}$ and $c_{{\rm A}_{k,2}}$, while the analogous
observables with flavour-dependent $\T$ angles, $\Phi_9^\qcd$ and $\Phi_{10}^\qcd$,
are sensitive to the entire set of parameters
$c_{{\rm A}_{k,1}}$, $c_{{\rm A}_{k,2}}$, $c_{{\rm A}_{k,3}}$ and $c_{{\rm A}_{k,4}}$.

The HQET expansions of our observables will be written
in terms of functions $\deta{i}(\Tq)$ and $\dphi{i}{j}(\Tq,\Th)$,
which are defined in appendix~\ref{sec:a1}, and we distinguish 
three cases:
\bi
\item[(i)] Observables depending on a single angle 
           (e.g., for $i=1$, 3, 6, and 11)
           \bd 
              \Phi^\hqet_i(\T_1) = \deta{i}(\T_1) + \sum_j \dphi{i}{j}(\T_1,\T_1) \cdot \omega_j
                          \equiv \Bigl( \deta{i} + \sum_j \dphi{i}{j}\cdot \omega_j\Bigr)\Big\vert_{\Tq=\Th=\T_1}
           \ed

\item[(ii)] Observables depending on two angles 
           (e.g., for $i=2$, 4, 7, and 8)
           \bea 
             \Phi^\hqet_i\Atwo
             & = & \deta{i}(\T_1) - \deta{i}(\T_2)
               + \sum_j\Big(\dphi{i}{j}(\T_1,\T_1) - \dphi{i}{j}(\T_2,\T_2)\Big) \cdot \omega_j 
             \nonumber\\
             & \equiv & \Big[ \deta{i} + \sum_j\dphi{i}{j} \cdot \omega_j\Big]^{\Tq=\Th=\T_1}_{\Tq=\Th=\T_2}
             \nonumber
           \eea

\item[(iii)] Observables depending on three angles 
           (e.g., for $i=5$, 9, and 10)
           \bea 
             \Phi^\hqet_i\Athree
             & = & \sum_j\Big(\dphi{i}{j}(\T_1,\T_2) - \dphi{i}{j}(\T_1,\T_3)\Big) \cdot \omega_j 
             \nonumber\\
             & \equiv & \Big[ \sum_j\dphi{i}{j} \cdot \omega_j\Big]^{\Tq=\T_1, \Th=\T_2}_{\Tq=\T_1, \Th=\T_3}
             \nonumber
           \eea
\ei
where any contribution from $\deta{i}(\Tq)$, which only depends on $\Tq$, can be
dropped because it cancels in the difference.

Observables of the form (i) are needed for the determination of HQET parameters, 
like $m_\bare$ or $Z_J$ factors, which determine the overall (re-)normalization 
of physical quantities. 
To illustrate the physical situations exploited by the observables of the 
form (ii) and (iii), we should keep in mind 
that in the SF the boundary fields create states which in the limit of large $T$ 
(and spatial volume) would include complicated multi-particle states 
(e.g., a B-meson plus a certain number of pions, including their excited states). 
Their total momentum is proportional to $\Th-\Tq$. Thus, observables of the 
form (ii) probe two different states with vanishing total momentum, while in 
(iii) one compares two states with also different total momentum (by giving 
different momenta to the heavy quark).

At tree-level and without background field all correlation functions 
with an insertion of $\Ospin$ vanish. In this case, none of the observables 
is sensitive to $\omegaspin$ (but it is also not needed to determine any of 
the other parameters). To ensure sensitivity to $\omegaspin$ at tree-level,
one could use a different setup with a non-trivial background field, similar 
to the one adopted in ref.~\cite{ospin:Nf0} for the computation of the 
renormalization constant of the $\Ospin$ operator. 
In refs.~\cite{HQET:param1m,HQET:Nf2param1m} it has  been shown that in any case at the
non-perturbative level a good sensitivity to $\omegaspin$ can be obtained also
with vanishing background field.

\subsection{HQET parameters of the Lagrangian and of the axial current}
We describe how the system of matching equations can be solved.
We use the shorthand notation introduced in the previous subsection and 
suppress the arguments $L$ and $a$. The explicit expressions for 
$\deta{i}(\Tq)$ and $\dphi{i}{j}(\Tq,\Th)$ are collected in 
appendix~\ref{sec:a1}. 

The observables are constructed in such a way that the bare mass enters only in 
the HQET expansion of 
\be
    \Phi^\hqet_1(\T_1) =   \big(
                   \deta{1} 
                 + \dphi{1}{1}\cdot m_\bare 
                 + \dphi{1}{2}\cdot\omegakin
                 + \dphi{1}{3}\cdot\omegaspin
                 + \dphi{1}{4}\cdot \ceff{A}{0}{1}
    \big)\Big\vert_{\Tq=\Th=\T_1}\;, 
   \label{e:mbare}
\ee
where we note that the contribution from $A_{0,1}$ would vanish if the matching were 
performed on a lattice of large temporal extent (as $\dphi{1}{4}(\Tq,\Th) \to 0$ for $T\to\infty$).

In order to solve the corresponding matching equation, one first needs to determine
$\omegakin$, $\omegaspin$, and $\ceff{A}{0}{1}$ from the matching equations 
for $\Phi_2,\cdots,\Phi_4$. First, 
\be
   \Phi^\hqet_2\Atwo =  \Big[ \deta{2} + \dphi{2}{2}\cdot\omegakin \Big]^{\Tq=\Th=\T_1}_{\Tq=\Th=\T_2}\;,
\label{e:Phi2}
\ee
and 
\be
   \Phi^\hqet_3(\T_1) = \left. \dphi{3}{3}\cdot\omegaspin \right\vert_{\Tq=\Th=\T_1}\;,
\label{e:Phi3}
\ee
allows determining $\omegakin$ and $\omegaspin$, respectively. Note that in the 
HQET expansion of $\Phi_2$ and $\Phi_3$ we have used the spin-symmetry relations 
$K_1^\stat = F_1^\stat$, $K_1^\kin = F_1^\kin$ and $K_1^\spin = -\frac{1}{3} F_1^\spin$.

Due to the choice $\Tq=\Th$, the HQET expansion 
\be
   \Phi^\hqet_4\Atwo = \Big[ 
                      \deta{4}
                    + \dphi{4}{2}\cdot\omegakin
                    + \dphi{4}{3}\cdot\omegaspin
                    + \dphi{4}{4}\cdot \ceff{A}{0}{1}
                    \Big]^{\Tq=\Th=\T_1}_{\Tq=\Th=\T_2}
\label{e:Phi4}
\ee
has no contribution from $A_{0,2}$. This permits us to extract $\ceff{A}{0}{1}$, 
and hence eq.~\ref{e:mbare} can be solved for $m_\bare$.

On the other hand, $\ceff{A}{0}{2}$ can be determined from matching $\Phi_5$,
which for $\T_1\not=\T_2$ or $\T_1\not=\T_3$ has the HQET expansion
\be
    \Phi^\hqet_5\Athree = \Big[
                      \dphi{5}{2}\cdot\omegakin
                    + \dphi{5}{4}\cdot \ceff{A}{0}{1}
                    + \dphi{5}{5}\cdot \ceff{A}{0}{2}
                    \Big]^{\Tq=\T_1, \Th=\T_2}_{\Tq=\T_1, \Th=\T_3}\;.
   \label{eq:phi5hqet}
\ee

Finally, $\ln Z^\hqet_{A_0}$ can be computed  (independently from $\Phi_5$) from 
matching $\Phi_6$, which has the HQET expansion
\be
   \Phi^\hqet_6(\T_1) = \left.\left(
                     \deta{6}
                   + \dphi{6}{2}\cdot\omegakin 
                   + \dphi{6}{3}\cdot\omegaspin
                   + \dphi{6}{4} \cdot \ceff{A}{0}{1}
                   + \ln Z^\hqet_{A_0} 
    \right)\right\vert_{\Th=\Tq=\T_1}\;.
\label{e:Phi6}
\ee
The $\T$ angles in the above equations can always be taken to be isotropic, 
i.e., $\T=(\theta,\theta,\theta)$. Moreover, we can use $\Th=\Tq$ except in 
$\Phi_5$, because otherwise $\dphi{5}{5} \sim f_{A_{0,2}}(\Tq,\Th)$ vanishes and hence 
there would be no  sensitivity to $\ceff{A}{0}{2}$.

%
The extension of the system of equations to include the matching of the spatial 
components of the axial current preserves the block structure
described in section~\ref{sec:s2}. In detail,
to determine $\ceff{A}{k}{1}$ and $\ceff{A}{k}{2}$, the matching equations 
for $\Phi_7$ and $\Phi_8$ need to be solved together. 
Their HQET expansion is
\be
    \Phi^\hqet_7\Atwo = \Big[ 
                  \deta{7}
                + \dphi{7}{2}\cdot\omegakin
                + \dphi{7}{3}\cdot\omegaspin
                + \dphi{7}{7}\cdot\ceff{A}{k}{1}
                + \dphi{7}{8}\cdot\ceff{A}{k}{2}
    \Big]^{\Tq=\Th=\T_1}_{\Tq=\Th=\T_2}\;,
\label{e:Phi7}
\ee
and
\be
    \Phi^\hqet_8\Atwo = \Big[ 
                  \deta{8} 
                + \dphi{8}{2}\cdot\omegakin
                + \dphi{8}{3}\cdot\omegaspin
                + \dphi{8}{7}\cdot\ceff{A}{k}{1}
                + \dphi{8}{8}\cdot\ceff{A}{k}{2}
    \Big]^{\Tq=\Th=\T_1}_{\Tq=\Th=\T_2}\;.
\label{e:Phi8}
\ee

Analogously, for $\ceff{A}{k}{3}$ and $\ceff{A}{k}{4}$, the matching equations 
for $\Phi_9$ and $\Phi_{10}$ have to be solved. 
Their HQET expansion reads
\be
    \Phi^\hqet_9\Athree = \Big[ 
                  \dphi{9}{2}\cdot\omegakin
                + \dphi{9}{7}\cdot\ceff{A}{k}{1}
                + \dphi{9}{8}\cdot\ceff{A}{k}{2}
                + \dphi{9}{9}\cdot\ceff{A}{k}{3}
                + \dphi{9}{10}\cdot\ceff{A}{k}{4}
    \Big]^{\Tq=\T_1,\Th=\T_2}_{\Tq=\T_1,\Th=\T_3}\;,
\label{e:Phi9}
\ee
and
\bea
    \Phi^\hqet_{10}\Athree &=& \Big[ 
                  \dphi{10}{2}\cdot\omegakin
                + \dphi{10}{7}\cdot\ceff{A}{k}{1}
                + \dphi{10}{8}\cdot\ceff{A}{k}{2} \nonumber \\ & &
                + \dphi{10}{9}\cdot\ceff{A}{k}{3}
                + \dphi{10}{10}\cdot\ceff{A}{k}{4}
    \Big]^{\Tq=\T_1,\Th=\T_2}_{\Tq=\T_1,\Th=\T_3} \;.
\label{e:Phi10}
\eea

Independently from $\Phi_9$ and $\Phi_{10}$, $\ln Z^\hqet_{\vec{A}}$ can be 
extracted from the matching of 
\bea
   \Phi^\hqet_{11}(\T_1) & = & \big(
                     \deta{11} 
                   + \dphi{11}{2}\cdot\omegakin 
                   + \dphi{11}{3}\cdot\omegaspin
   \nonumber\\ & &
                   + \dphi{11}{7} \cdot\ceff{A}{k}{1}
                   + \dphi{11}{8} \cdot\ceff{A}{k}{2} 
                   + \ln Z^\hqet_{\vec{A}} 
   \big)\Big\vert_{\Tq=\T_{\ell'}=\Th=\T_1}\;,
\label{e:Phi11}
\eea
which uses a three-point function. 
Alternatively, one can take the observable as in eq.~\ref{e:Phi11old} 
\bea
   \Phi^{\prime\,\hqet}_{11}(\T_1) & = & \big(
                     \deta{11}^\prime 
                   + \dphi{11}{\prime\,2}\cdot\omegakin 
                   + \dphi{11}{\prime\,3}\cdot\omegaspin
   \nonumber\\ & &
                   + \dphi{11}{\prime\,7} \cdot\ceff{A}{k}{1}
                   + \dphi{11}{\prime\,8} \cdot\ceff{A}{k}{2} 
                   + \ln Z^\hqet_{\vec{A}} 
   \big)\Big\vert_{\Tq=\Th=\T_1}\;,
\label{e:Phi112pt}
\eea
which is defined through two-point functions only.

\section{Tree-level results for heavy-light currents at $\Order{1/m_{\rm h}}$}
\label{sec:s4}
We perform the tree-level computation using $\Order{a}$ improved Wilson quarks. For the static
action, at this order, the Eichten-Hill discretization in ref.~\cite{stat:eichhill1} and the HYP1/2
 actions in ref.~\cite{HQET:statprec} coincide. 
We numerically evaluate the tree-level expressions for the relevant correlation functions
relying on the known  formulae for the relativistic and static SF-propagators. 
Those can be found in refs.~\cite{impr:pap2,zastat:pap1}.
In QCD we fix $z=\left\{4,8,12,16,20,24,28,32,64\right\}$ with $z=\tilde{m}_{\rm q}L$ and
\be
\tilde{m}_{\rm q}=m_{\rm q} (1+b_{\rm m}am_{\rm q})\;,\quad am_{\rm q}=\frac{1}{2}\left(\frac{1}{\kappa}-\frac{1}{\kappa_{\rm c}}\right)\;,
\ee
where $\kappa_{\rm c}$ is the critical value of the Wilson hopping-parameter
and $b_{\rm m}$ is an improvement coefficient~\cite{impr:pap1}. At tree-level $\kappa_{\rm c}=1/8$ and
$b_{\rm m}=-1/2$. Of course, at this order other definitions of the heavy-quark mass could have been used (see, e.g., the one adopted in appendix~D of ref.~\cite{HQET:param1m}). The one we chose is the simplest consistent with $\Order{a}$ improvement.

At tree-level we can take the continuum limit of each individual term in the HQET expansion.
We do that by a linear extrapolation in $a^2$ of results obtained for $128\leq L/a \leq 256$.
We adopt the same procedure for the quantities computed in  QCD  at each value of $z$.
These extrapolations are always well controlled, however, given the high precision of the available data,
the growth of cutoff effects with $z$ is clearly visible in the QCD results; 
we provide an example in fig.~\ref{fig:phi4cutoff}. 
Therefore, $a^3$-terms as well as $a^4$-terms had sometimes to be included in 
the fits.
\begin{figure}[h!]
\begin{center}
\includegraphics[width=12.9cm]{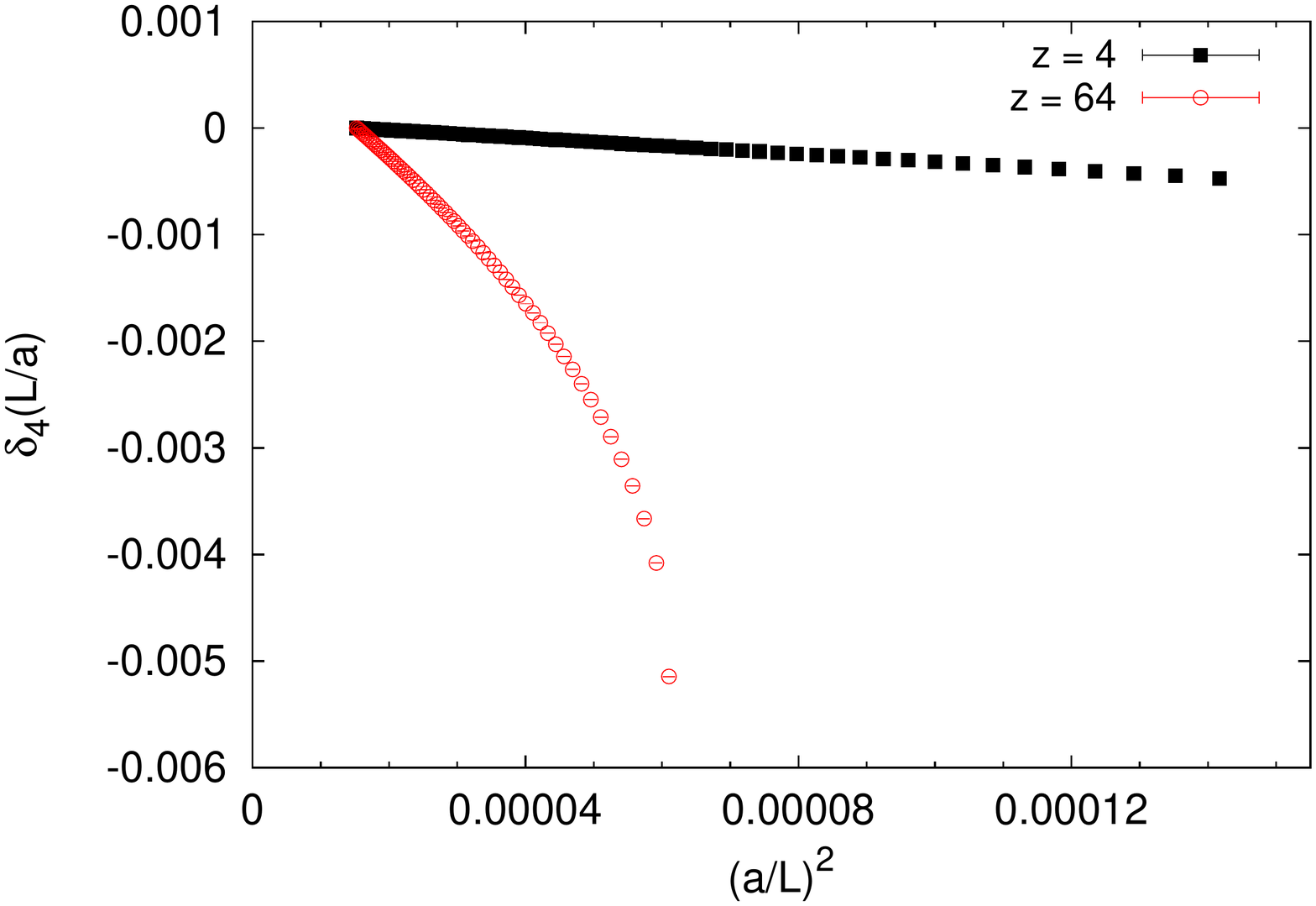} 
\caption{Tree-level results for $\delta_4(L/a)$  defined as the 
difference between $\Phi_4^\qcd(1, 0.5)$ at a given value of $L/a$ and $\Phi_4^\qcd(1, 0.5)$ at $L/a=256$,
for $z=4$ and $z=64$.}
\label{fig:phi4cutoff}
\end{center}
\end{figure}

Once we have performed the continuum limit extrapolations for all the 
$\Phi_i^{\rm QCD}$, $\eta_i$, and $\varphi_i^j$ quantities, 
we solve the system of matching equations \ref{e:phiexp}
for the unknown HQET parameters
\be
    \omega_k = \left(\varphi^{-1}\right)_k^i \cdot \left(\Phi_i^{\rm QCD} -\eta_i\right) \;.
\ee
Due to the block structure of the matrix $\varphi$ and 
to the additional simplifications arising at tree-level, the parameters 
can be determined one after the other.
For instance, making use of tree-level relations as those 
in appendix~D of ref.~\cite{HQET:param1m}, one finds that
$\dphi{11}{2}$ (using either three- or two-point functions)
and $\dphi{6}{2}$ vanish in the continuum. 
We also recall that all the terms proportional to $\omega_{\rm spin}$ 
vanish at tree-level.

As an example we consider $\Phi_4^{\rm QCD}$. In order to determine 
$\ceff{A}{0}{1}$ from
\be
\label{e:cAeff}
\frac{\left(\Phi_4^\qcd(L,\tilde{m}_{\rm q},0)-\eta_4(L,0)
-\varphi_4^2(L,0)\,\omegakin
\right)}
{\varphi_4^4(L,0)} = \ceff{A}{0}{1}\;,
\ee
one has to first determine other parameters, namely $\omega_{\kin}$ 
in this case~\footnote{A term $-\varphi_4^3(L,0)\,\omegaspin$ in the numerator on the l.h.s. of eq.~\ref{e:cAeff} has been dropped as it
vanishes at tree-level.}. Then, with $\ceff{A}{0}{1}$, one can determine 
$m_\bare$, $\ceff{A}{0}{2}$, and $\ln Z^\hqet_{A_0}$, etc.

In order to investigate the size of higher-order contributions 
in $\mhinv$, we multiply each parameter with appropriate
factors of $\mh\equiv\tilde{m}_{\rm q}$ or $L$. Then,
for $z\equiv \mh\cdot L\to \infty$, these combinations approach the known classical 
values, for example
\bea
  \ceff{A}{0}{1} \cdot \mh & = & -\frac{1}{2} + \Order{1/z} \;,\\
   \ln Z^\hqet_{A_0} & = & 0 + \Order{1/z} \;, \\
m_{\rm bare}-m_{\rm h} & = & 0 + \Order{1/z} \;.
\eea
Note that in the case of  $\omega_1 \equiv m_\bare$
both, the term of order $\mh$ and the term of order $1$ (which vanishes at tree level),
are already fixed by matching in the static approximation.
Any $\Order{1/z}$ deviations from the above behaviour are then a sign of the 
$\mhinv$-corrections neglected in the static theory.

If all $\mhinv$-terms are included in HQET and in the matching with QCD,
as done throughout this paper, then also the $\Order{1/z}$-corrections to the parameters 
are fixed. The deviation from a linear $1/z$-dependence is then a sign of 
higher-order corrections neglected in the effective theory. These corrections
also give rise to a non-vanishing dependence of the parameters on the
specific $\T$ angles in the choice of the observables used in the matching.

When solving the  system of matching equations by backward substitution,
the higher-order corrections in $1/z$ enter through the non-linear $1/z$-dependence 
of the already determined parameters as well as through the higher-order 
$1/z$-dependence of the QCD observable itself. To disentangle these 
two sources we solve each equation in two ways: either we directly use
the results of the previously determined parameters at the same value 
of $z$ and for the choice of $\T$ angles, which gave the smallest 
higher-order corrections in $1/z$, or we use their classical values, 
e.g., $-1/(2m_{\rm h})$ for $\omega_{\kin}$ in the equation for $\ceff{A}{0}{1}$. 
The $1/z$-dependence of the parameters obtained in both ways is shown
in the figures~\ref{fig:Lplots} -- \ref{fig:Vkplots}. The values 
obtained in the latter way 
are labeled with a superscript ``EX'', indicating 
that in this case the $1/z$-corrections are {\em exclusively} due to the QCD 
observable and not inherited from previously determined parameters.

The parameter $\omega_{\rm kin}$ enters in the matching equations for
many of the other parameters. From fig.~\ref{fig:Lplots} we see that 
$\omega_\kin$ has large $1/z$-corrections, although they are to a very
good approximation linear in $1/z$ and $\T$-independent. Propagation of 
these large effects into other parameters, where $\omega_\kin$ enters 
in the matching equation, is then suppressed in the ``EX'' setup.

Although it is rather difficult to single out a unique choice for the $\T$
angles producing the smallest higher-order corrections across all parameters, we 
collect in table~\ref{t:angles} those values which produced the smallest higher-order 
corrections for each parameter (and for $z \ge 10$).
The selection was performed without the ``EX'' option in order to stay in a 
situation which is closer to the one encountered when performing non-perturbative 
matching.
Of course, our choice has been made only from a small set of
$\T$ angles (in a rather large range), thus it can only give 
an rough indication of the optimal values.

\begin{table}[htb]
\def\different{{$(\ast)$}}
\begin{center}
\renewcommand{\arraystretch}{1.25}
\begin{tabular}{@{\extracolsep{0.2cm}}lccc}
\toprule
parameter	   &  $\T_1$	&   $\T_2$               &  $\T_3$ \\
\midrule
\hline
$m_\bare$	   &	0.0	&   ---	                 &  ---      \\
$\omegakin$	   &  0.0 - 0.5	&  0.5 - 1.0  \different &  ---  \\
\hline
$\ceff{A}{0}{1}$   &    0.0     &     0.5                & ---  \\
$\ceff{A}{0}{2}$   &    0.5     &     0.0                &  1.0 \\
$\ln Z_{A_0}$       &    0.0     &    ---                 & ---  \\
$\ceff{A}{k}{1}$   &  0.5 - 1.5 &  0.5 - 1.0  \different & --- \\
$\ceff{A}{k}{2}$   &  0.5 - 1.5 &  0.5 - 1.5  \different & --- \\
$\ceff{A}{k}{3}$   &    0.5     &      1.5                & 1.0 \\
$\ceff{A}{k}{4}$   &    0.5     &      1.0                & 0.0 \\
$\ln Z_{A_k}$       &    0.0     &      ---               & --- \\
\hline
$\ceff{V}{0}{1}$   &  0.5 - 1.5 &   0.5 - 1.5 \different & --- \\
$\ceff{V}{0}{2}$   &    0.5     &      0.5               & 1.0 \\
$\ln Z_{V_0}$       &    0.0     &      ---               & --- \\
$\ceff{V}{k}{1}$   & 0.0,0.5,0.5 & 0.0,1.0,1.0           & --- \\
$\ceff{V}{k}{2}$   &     1.0     &     1.5               & --- \\
$\ceff{V}{k}{3}$   & 0.0,1.5,1.5 & 0.0,0.5,0.5           & 0.0,1.0,1.0 \\
$\ceff{V}{k}{4}$   &     1.5     &     0.5               & 1.0 \\
$\ln Z_{V_k}$	   &     0.0     &     ---               & --- \\
\bottomrule
\end{tabular}
\end{center}
\caption{Our preferred choice of $\T$ angles to reduce higher-order corrections in the parameters. 
    The \different\ indicates that $\T_1\not=\T_2$ is required.
The shorthand $\vec{\theta}=t$ is used for $\vec{\theta}=(t,t,t)$.
}
\label{t:angles}
\end{table}
From the lower panels of figs.~\ref{fig:Lplots} -- \ref{fig:Vkplots},
we note that the choice $\T_1=0$ is clearly the best for 
$m_{\rm bare}$, $\ln\,Z_{A_0}$, $\ln\,Z_{\vec{\rm A}}$, $\ln\,Z_{V_0}$ 
and $\ln\,Z_{\vec{\rm V}}$. 
These are the parameters determined from observables which 
depend only on a {\em single} angle $\T_1$ (corresponding 
to case (i) of section \ref{subsec:obs}). At tree-level the dependence
of these matching equations on any other parameter vanishes for $\T_1=0$.
As a consequence, these parameters do not depend on whether one uses 
the ``EX'' setup or not, and the higher-order corrections seem
to vanish.
On the other hand, already at $\T=\vec{1}$ the corrections are 
very significant, especially in the static approximation.

For the parameters determined from observables which depend on two angles
(corresponding to case (ii) of section \ref{subsec:obs}), like
$\ceff{A}{0}{1}$, $\ceff{A}{k}{1}$, $\ceff{A}{k}{2}$, $\ceff{V}{0}{1}$, 
$\ceff{V}{k}{1}$, and $\ceff{V}{k}{2}$, the dependence on the $\T$ angles 
is moderate, see upper panels of figs. \ref{fig:A0plots} -- \ref{fig:Vkplots}. 
Smaller values of $(\T_1,\T_2)$ often yield smaller or less non-linear 
$1/z$-corrections.
Moreover, some of the higher-order $1/z$-dependence is inherited from the 
other parameters because it tends to be smaller in the ``EX'' setup.

The parameters determined from observables which depend on three angles
typically can depend significantly on the choice of these angles, and 
the effect of higher-order corrections in $1/z$ is sizable.
These effects do not decrease strongly (if at all) in
the ``EX'' case. Therefore they are a genuine property of these 
observables rather than inherited from other parameters. 

Finally, in fig.~\ref{fig:V0-2and3} we compare the two determinations of 
$\ln\,Z_{{\rm V}_0}^\hqet$, using either two- or three-point functions.
Because of the significantly flatter $1/z$-dependence, the latter is
clearly preferable. The analogous conclusion is found for $\ln\,Z_{\vec{\rm A}}^\hqet$.

For the parameters that enter in the HQET expansion of $A_0$ and $V_k$,
the higher-order corrections seem to be inherited to a large extent from $\omega_{\rm kin}$.
Since $A_0$ and $V_k$ are the current components
relevant for the computations of the ${\rm B}_{({\rm s})}$-meson decay constant and the form 
factors of semileptonic decays to light mesons, it would be important to find a way to
improve the corresponding matching conditions and the one for $\omega_{\rm kin}$.
We describe a possible modification of all the observables to be used
in the non-perturbative matching, which exploits the tree-level results
in order to reduce the higher-order effects in $1/z$.
We consider eq.~\ref{e:phiexp} at tree-level and replace all the parameters
$\omega_i$ by their known classical values $\omega_i^{\rm cl}$. 
This allows us to define a set of coefficients $\delta_{\Phi_i}^{\rm tree}(L,M)$ 
through
\be
\Phi_i^{\qcd,\rm tree}(L,M,0)=
\left(\eta_i^{\rm tree}(L,0)
+\varphi_i^{j,\rm tree}(L,0)\,\omega_j^{\rm cl}(M)\right) 
\times \left(1+\delta_{\Phi_i}^{\rm tree}(L,M)\right)\;. 
\ee
Clearly, we are free to modify the matching condition by higher-order terms, 
e.g., $\Order{1/z^2}$. In particular, we may use
$\Phi_i^{\qcd}(L,M,0)/(1+\delta_{\Phi_i}^{\rm tree}(L,M))$
in the non-perturbative matching. In this way, higher-order effects in $1/z$ are
completely removed at tree-level. The coefficients $\delta_{\Phi_i}^{\rm tree}(L,M)$ are a
rather simple byproduct of
the results presented here and will be published together with the non-perturbative study.
\begin{figure}[h!]
\begin{center}$
\begin{array}{c}
\includegraphics[width=9.75cm]{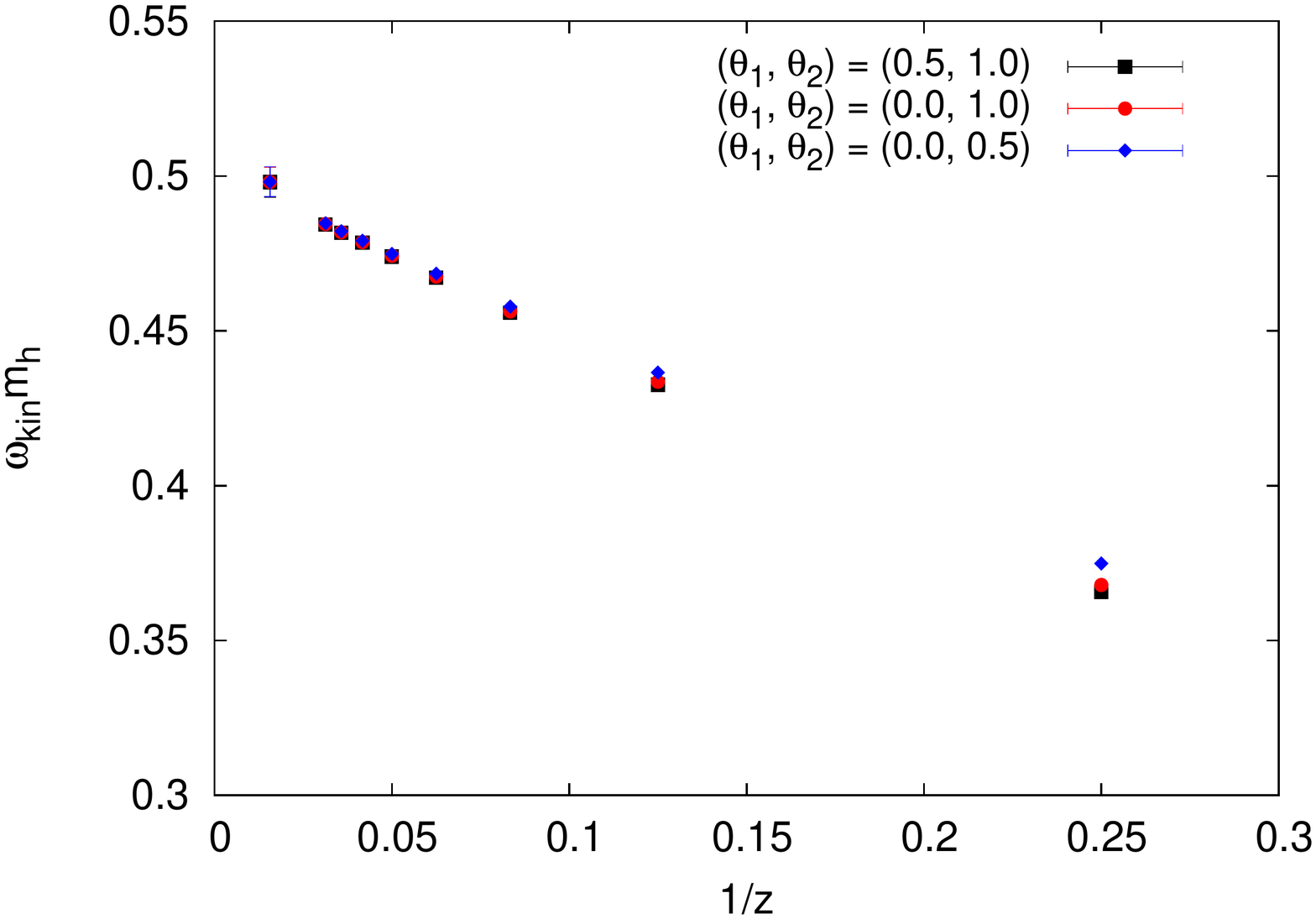} \\
\includegraphics[width=9.75cm]{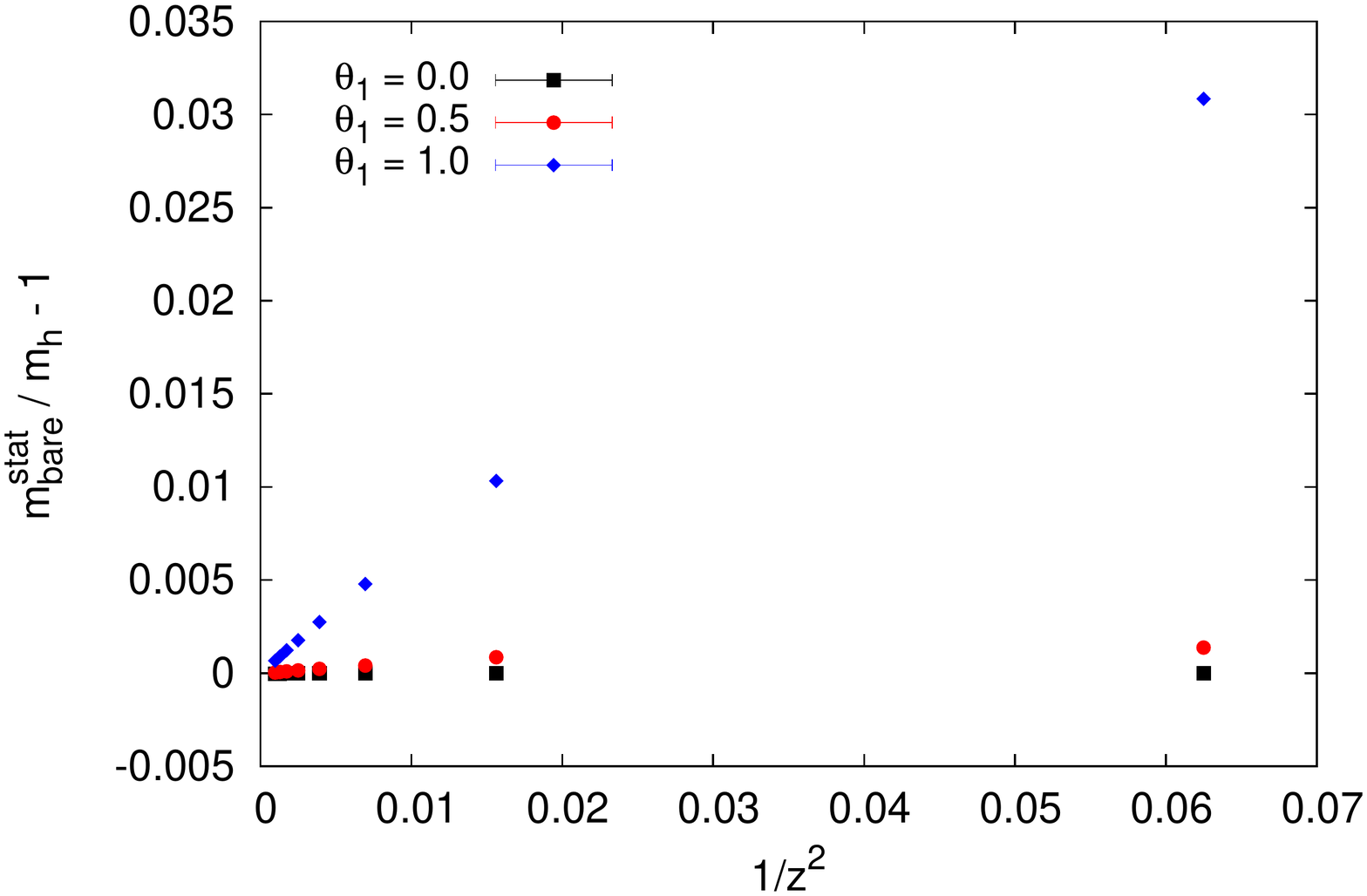} \\
\includegraphics[width=9.75cm]{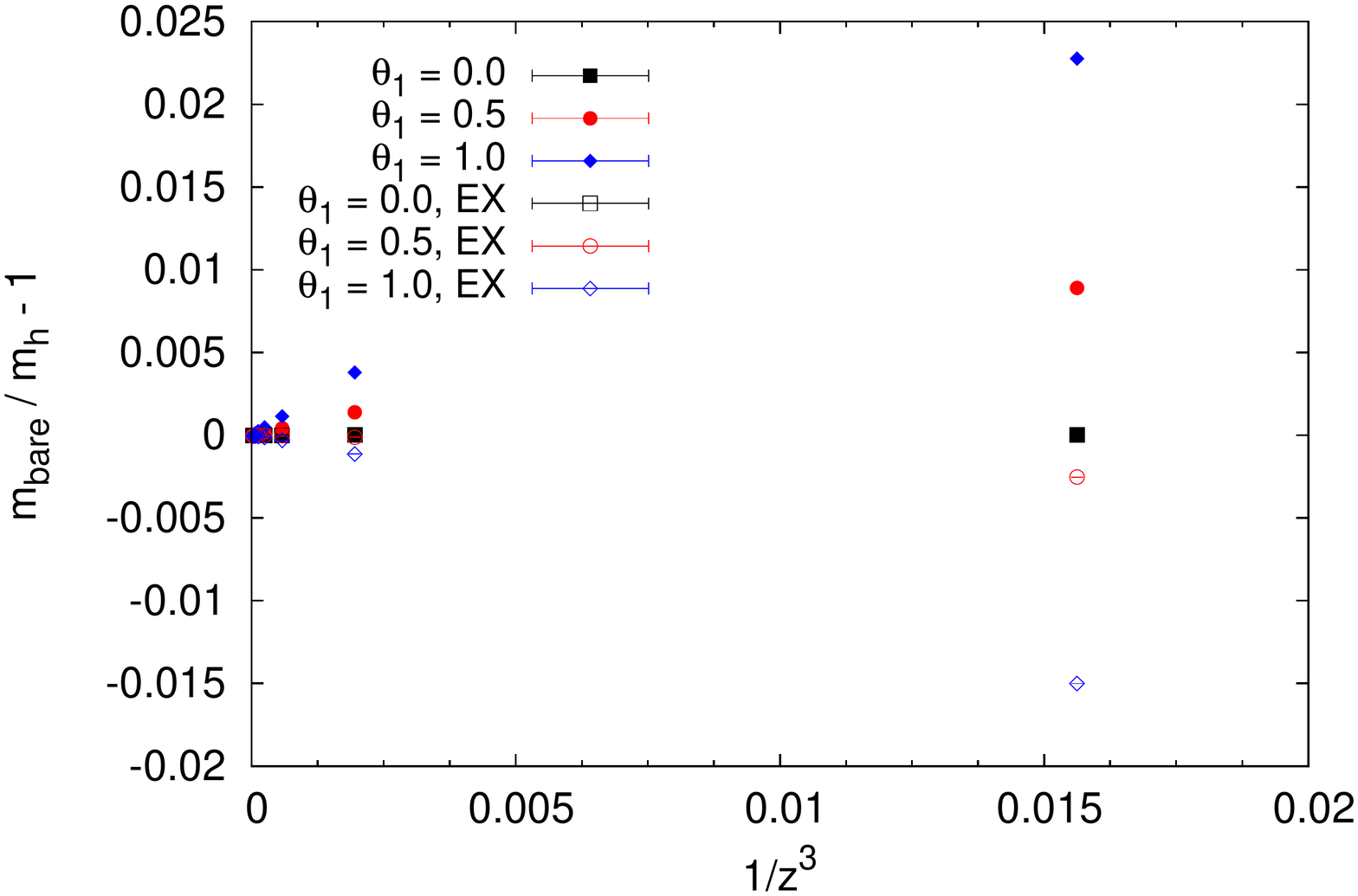}
\end{array}$
\end{center}
\caption{Tree-level continuum results for the parameters in $\lag{HQET}$.}
\label{fig:Lplots}
\end{figure}
\begin{figure}[h!]
\hspace{-2.cm}
$\begin{array}{cc}
\includegraphics[width=9.75cm]{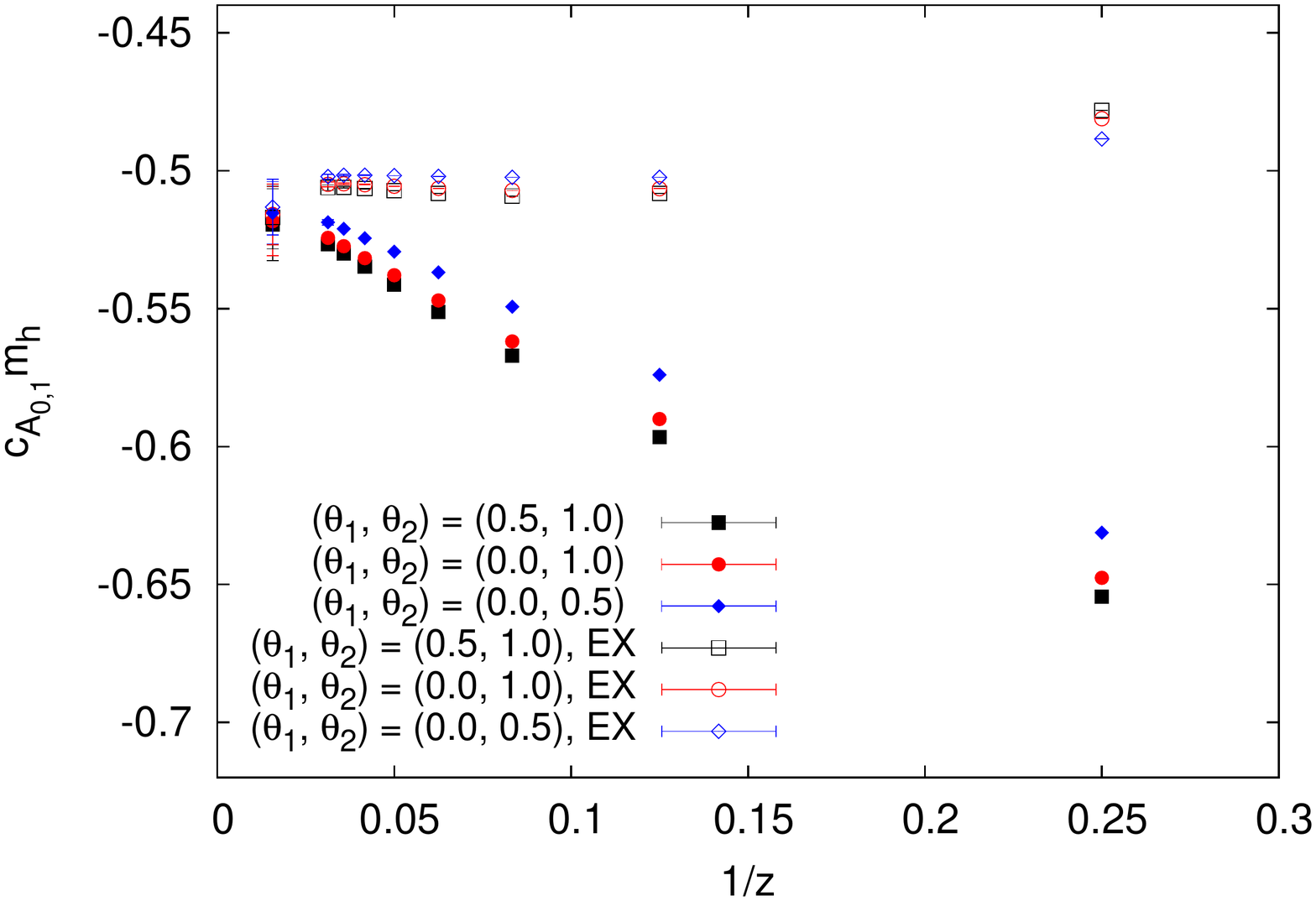} & 
\hspace{-1.8cm}\includegraphics[width=9.75cm]{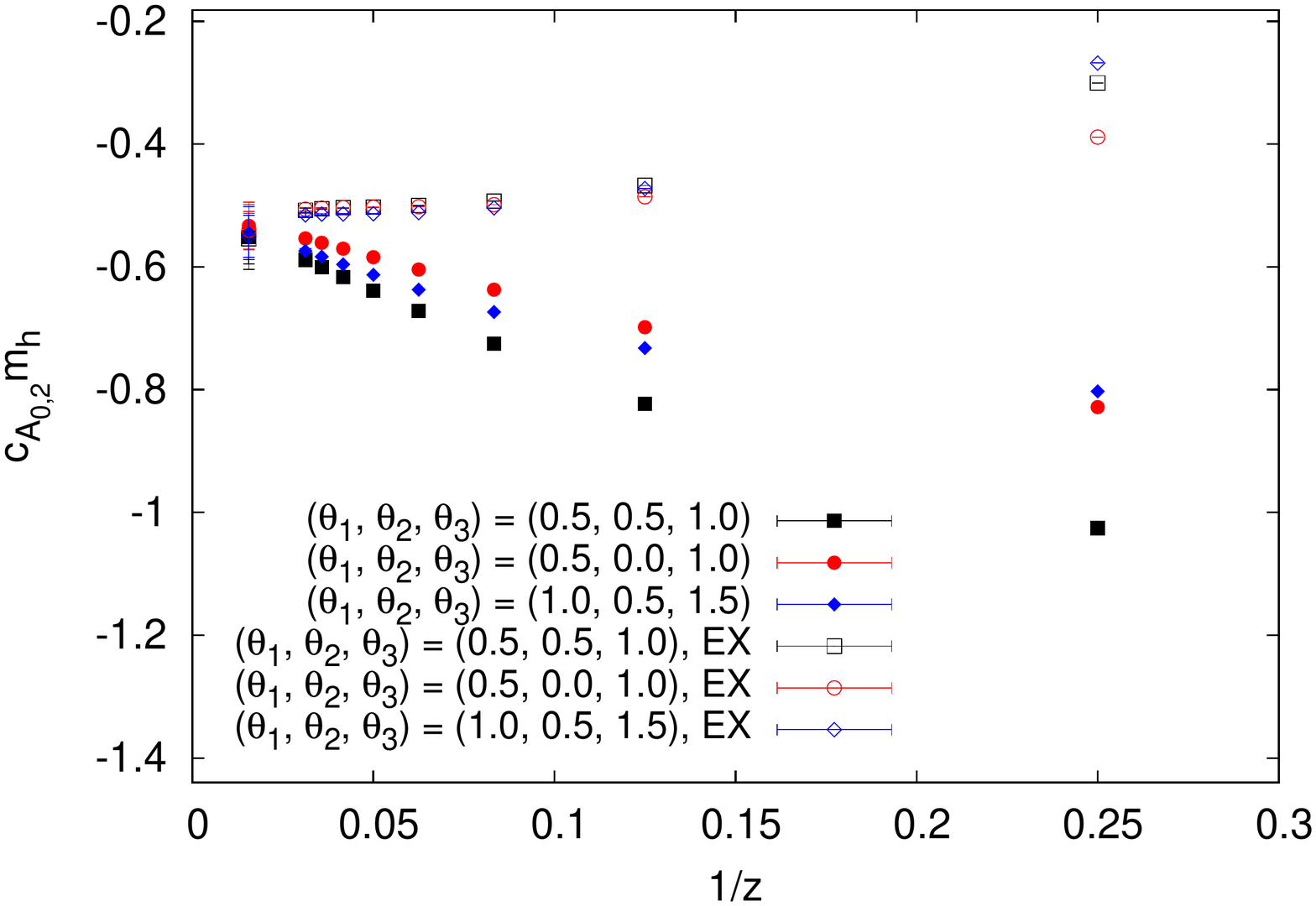}
\\
\includegraphics[width=9.75cm]{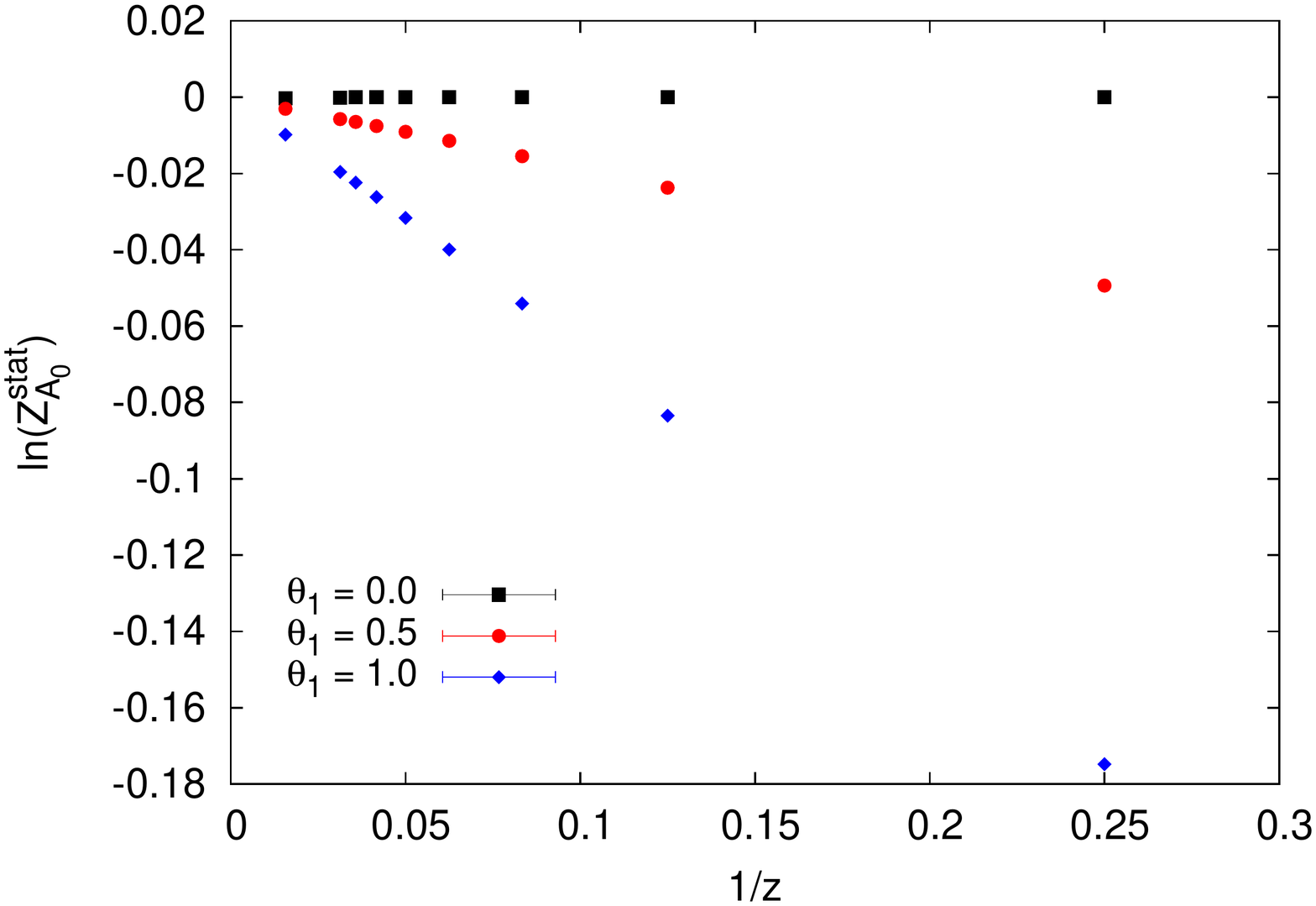} &
\hspace{-1.8cm}\includegraphics[width=9.75cm,height=6.68cm]{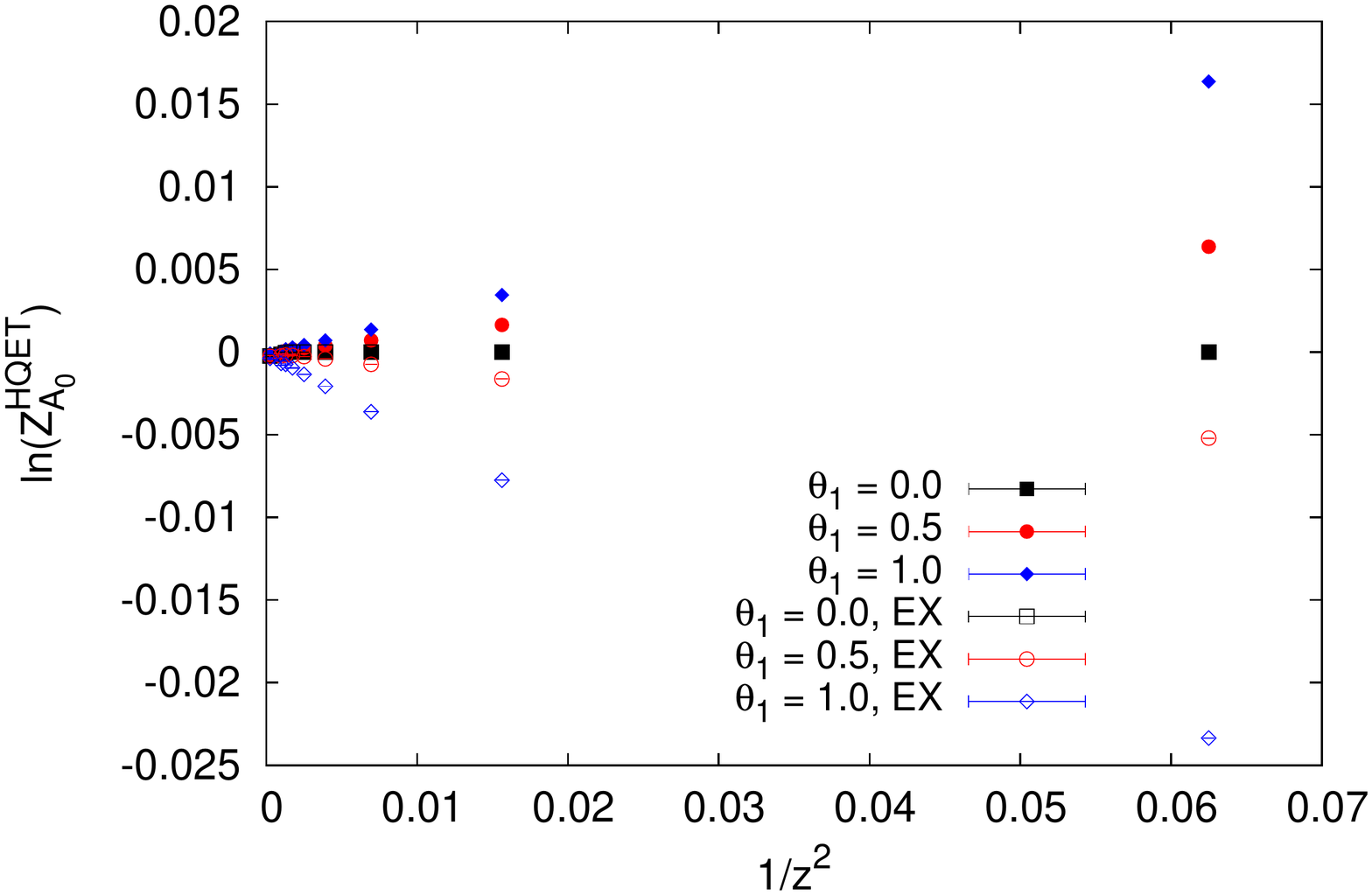}
\end{array}$
\caption{Tree-level continuum results for the parameters of the temporal component of the axial current.}
\label{fig:A0plots}
\end{figure}
\begin{figure}[h!]
\hspace{-2.cm}
$\begin{array}{cc}
\includegraphics[width=9.75cm]{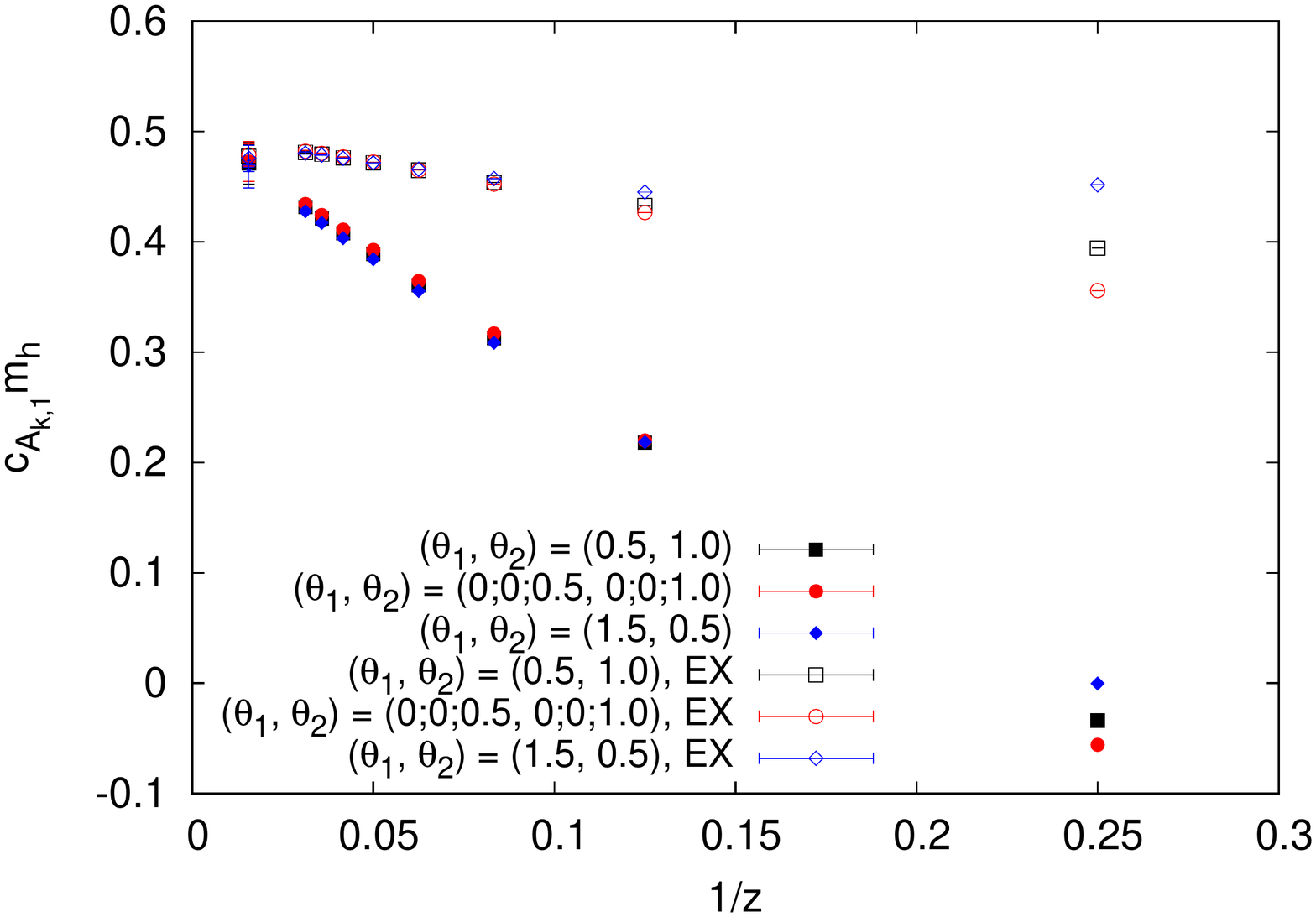} & 
\hspace{-1.8cm}\includegraphics[width=9.75cm]{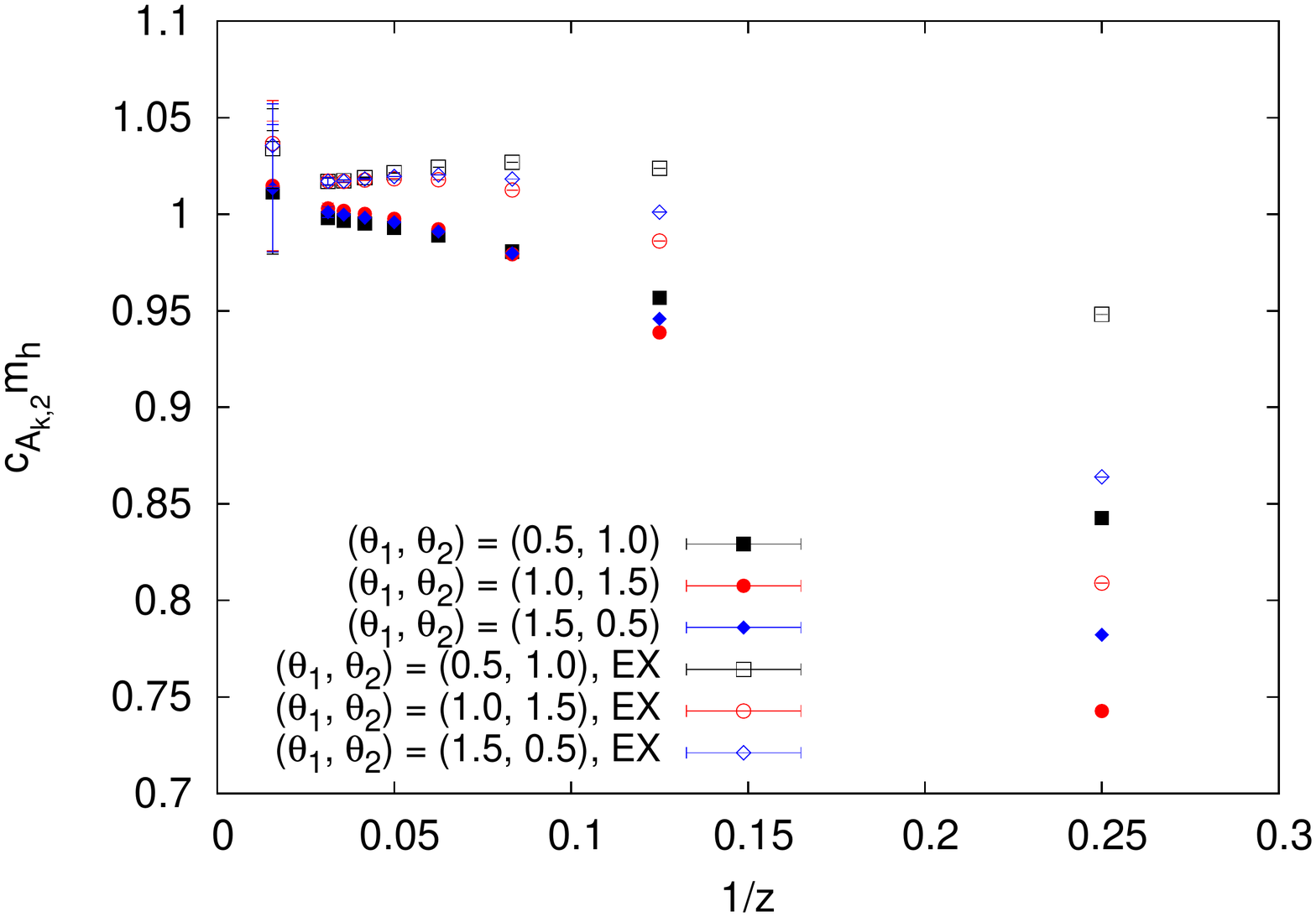} \\
\includegraphics[width=9.75cm]{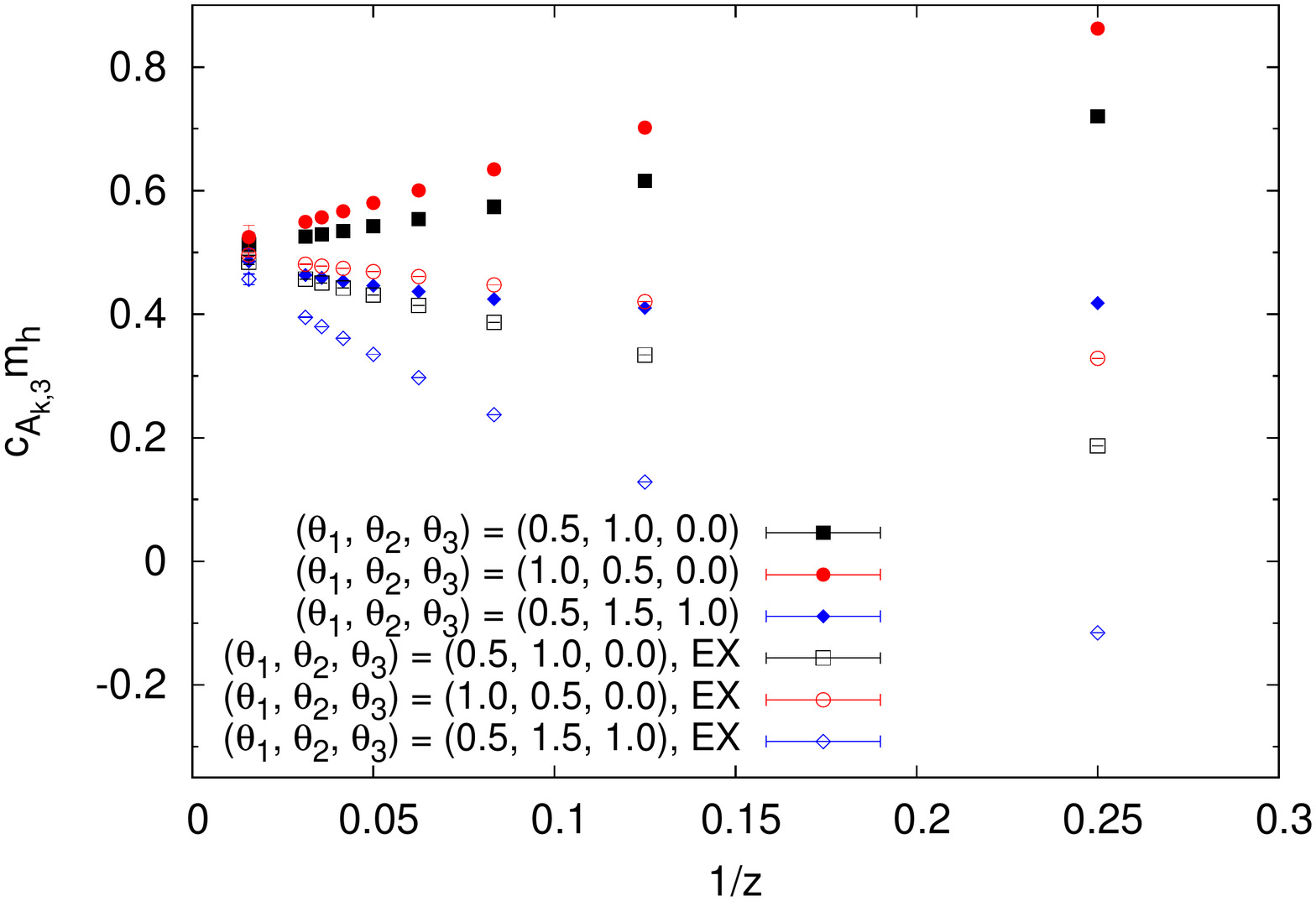} & 
\hspace{-1.8cm}\includegraphics[width=9.75cm]{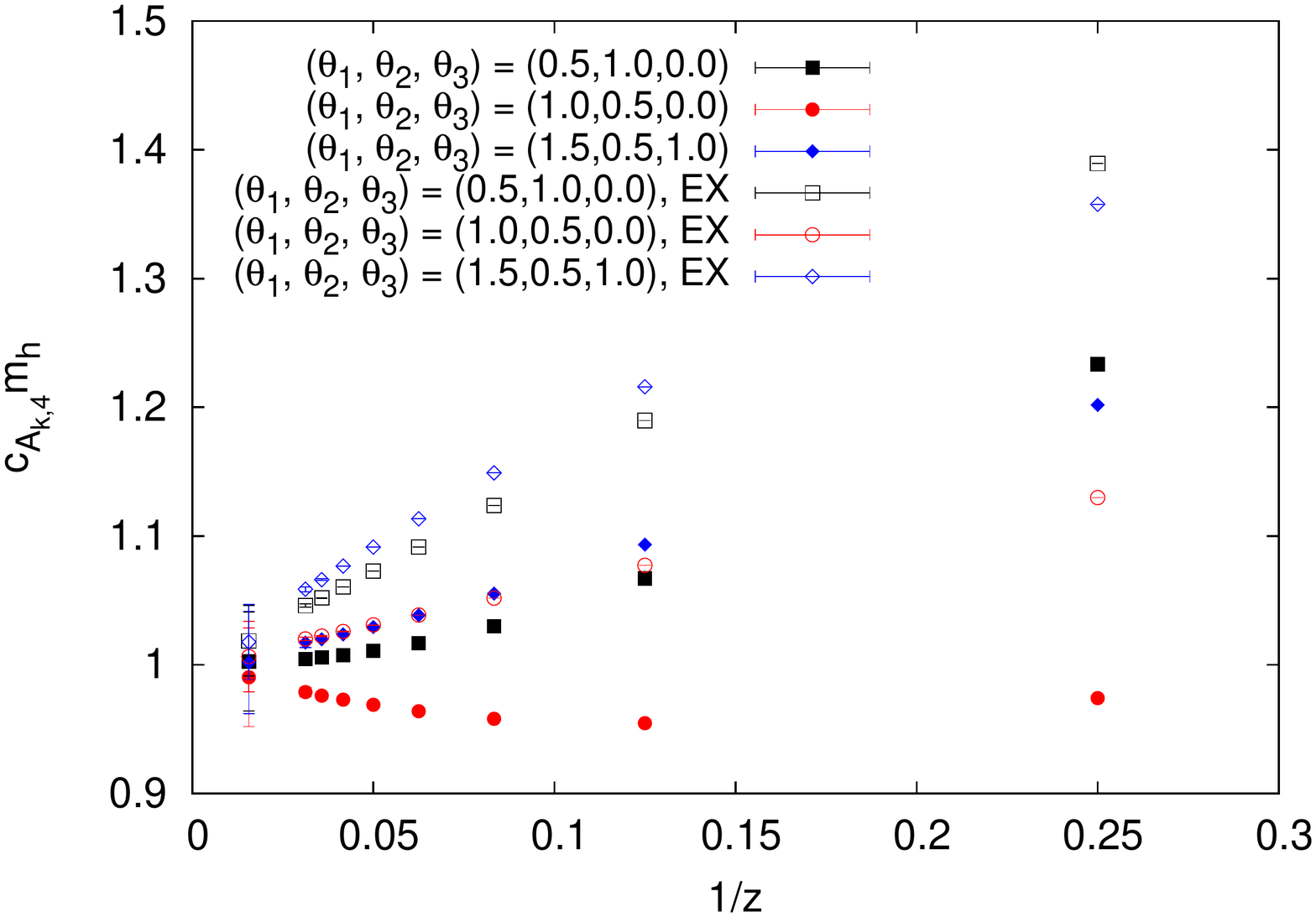} \\
\includegraphics[width=9.75cm]{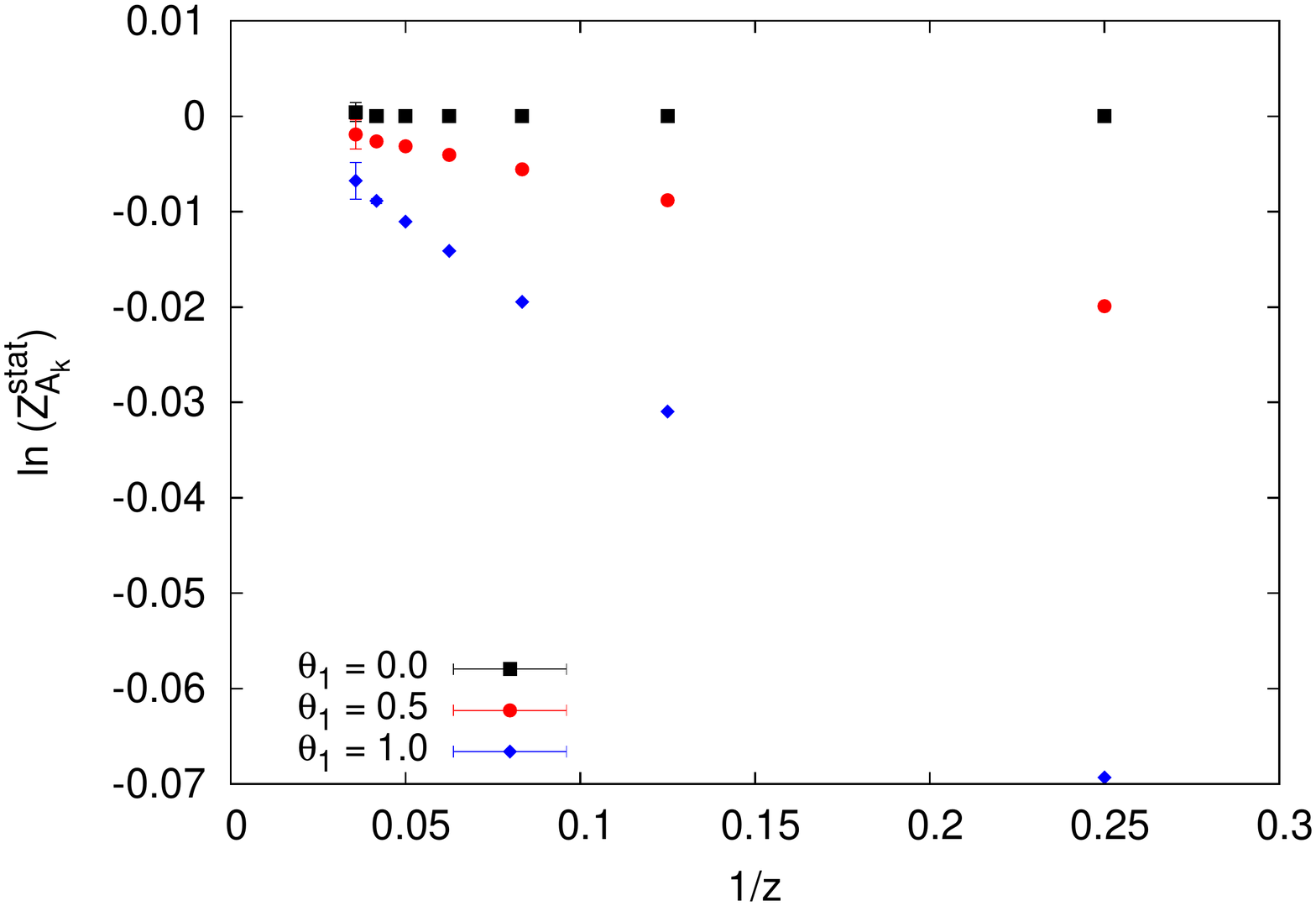} &
\hspace{-1.8cm}\includegraphics[width=9.75cm,height=6.68cm]{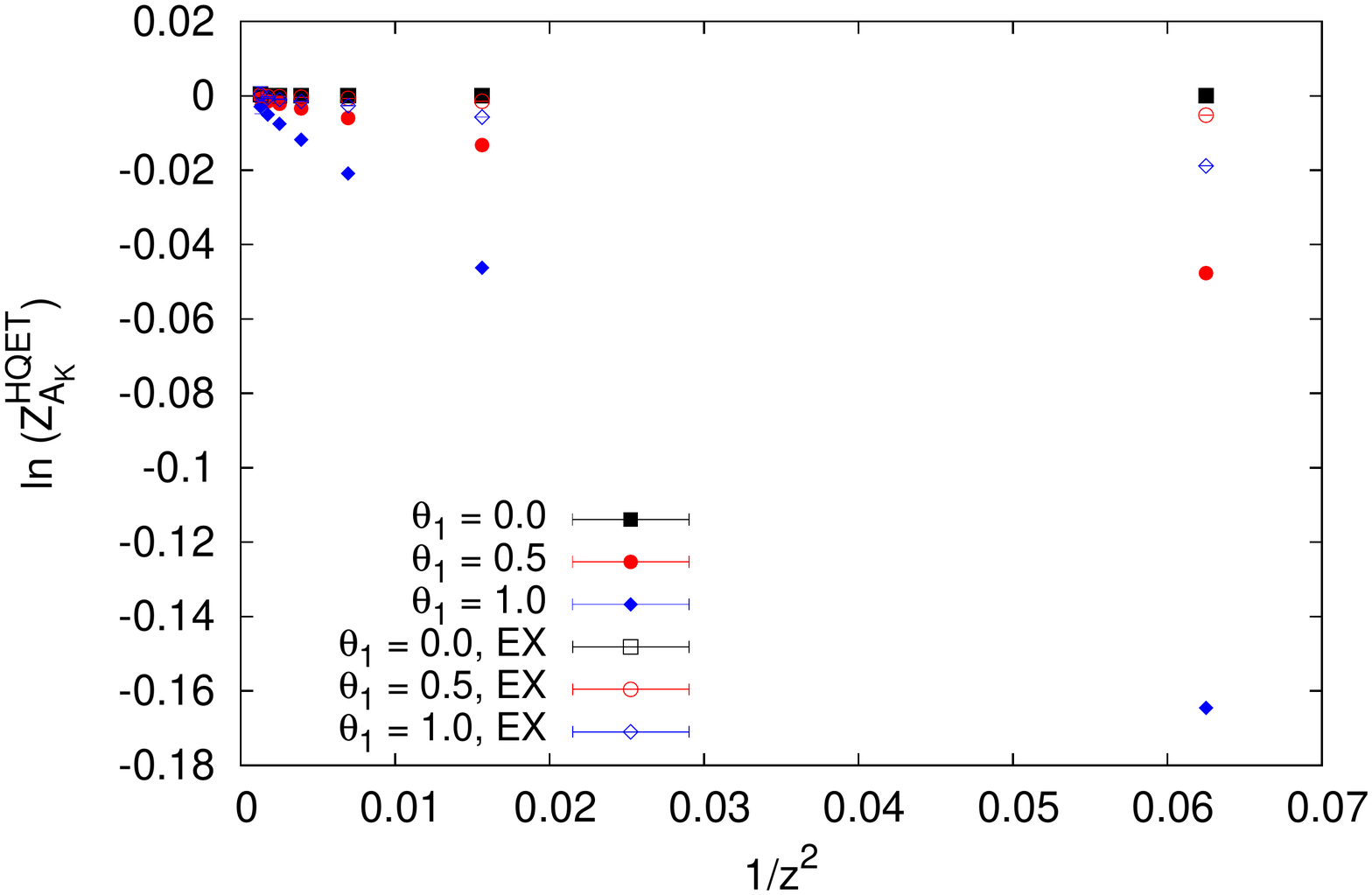}
\end{array}$
\label{fig:Akplots}
\caption{Tree-level continuum results for the parameters of the spatial components of the axial current.}
\end{figure}
%
\begin{figure}[h!]
\hspace{-2.cm}
$\begin{array}{cc}
\includegraphics[width=9.75cm]{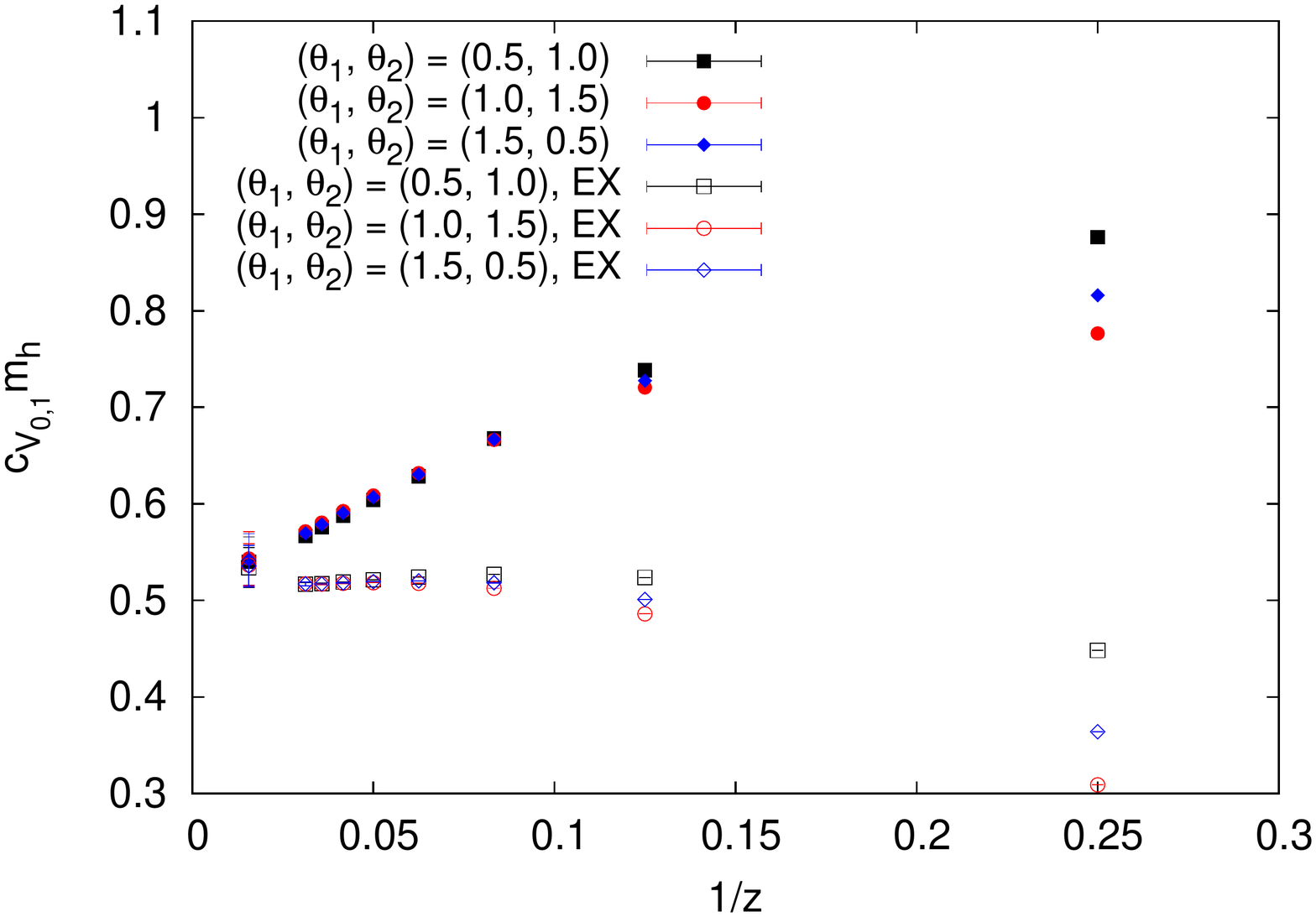} & 
\hspace{-1.8cm}\includegraphics[width=9.75cm]{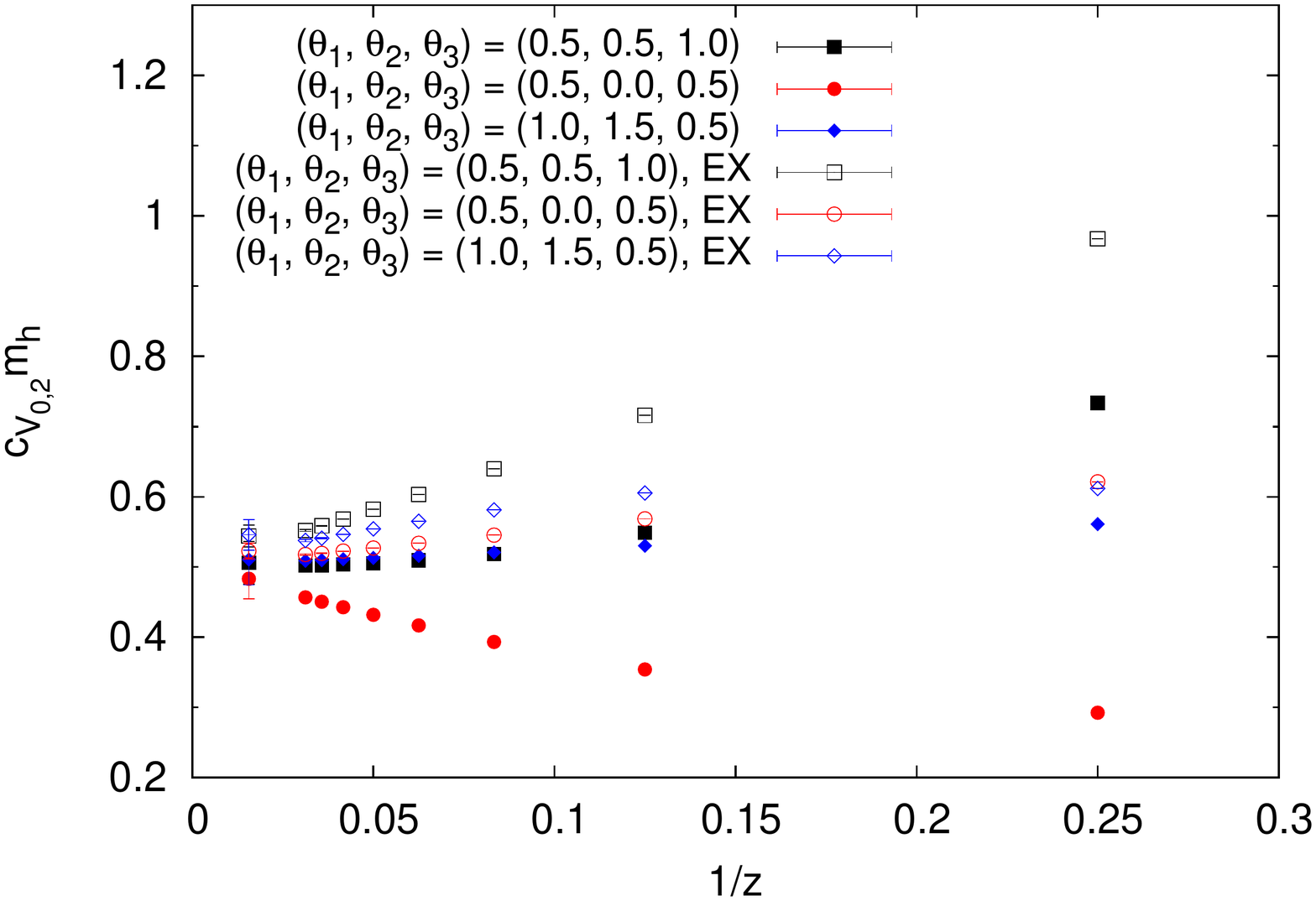}
\\
\includegraphics[width=9.75cm]{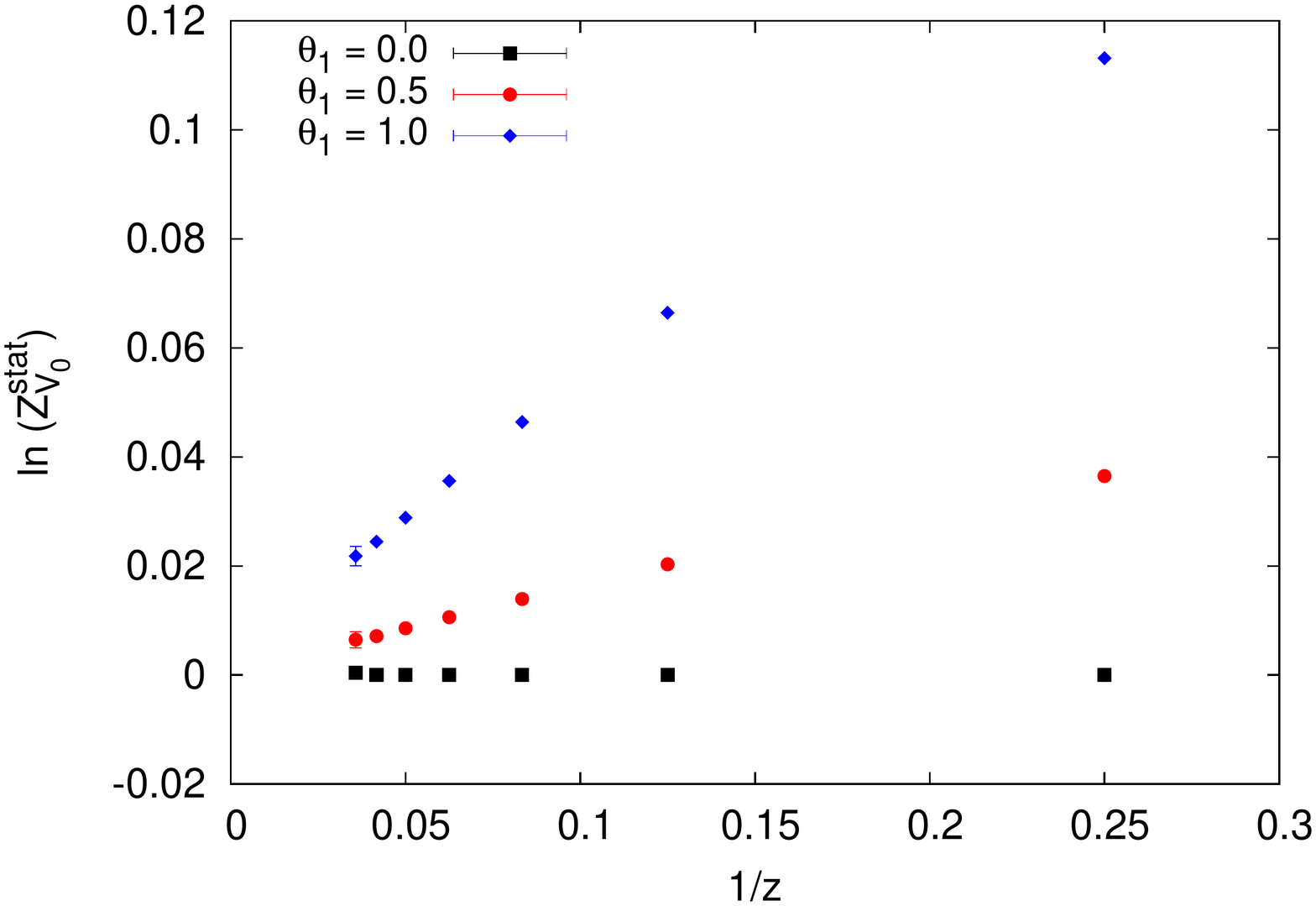} &
\hspace{-1.8cm}\includegraphics[width=9.75cm,height=6.68cm]{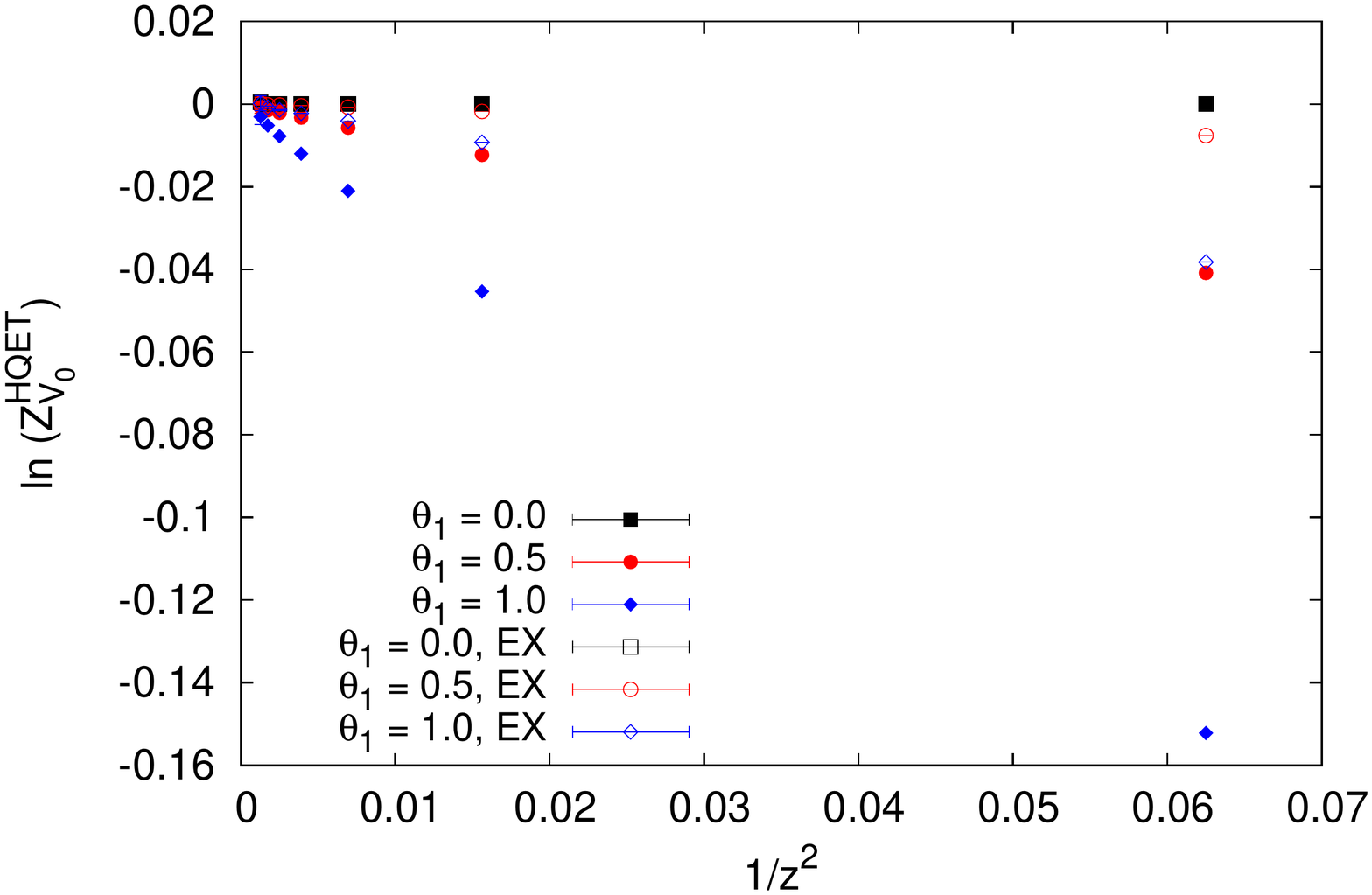}
\end{array}$
\caption{Tree-level continuum results for the
parameters of the temporal component of the vector current.}
\label{fig:V0plots}
\end{figure}
\begin{figure}[h!]
\hspace{-2.cm}
$\begin{array}{cc}
\includegraphics[width=9.75cm]{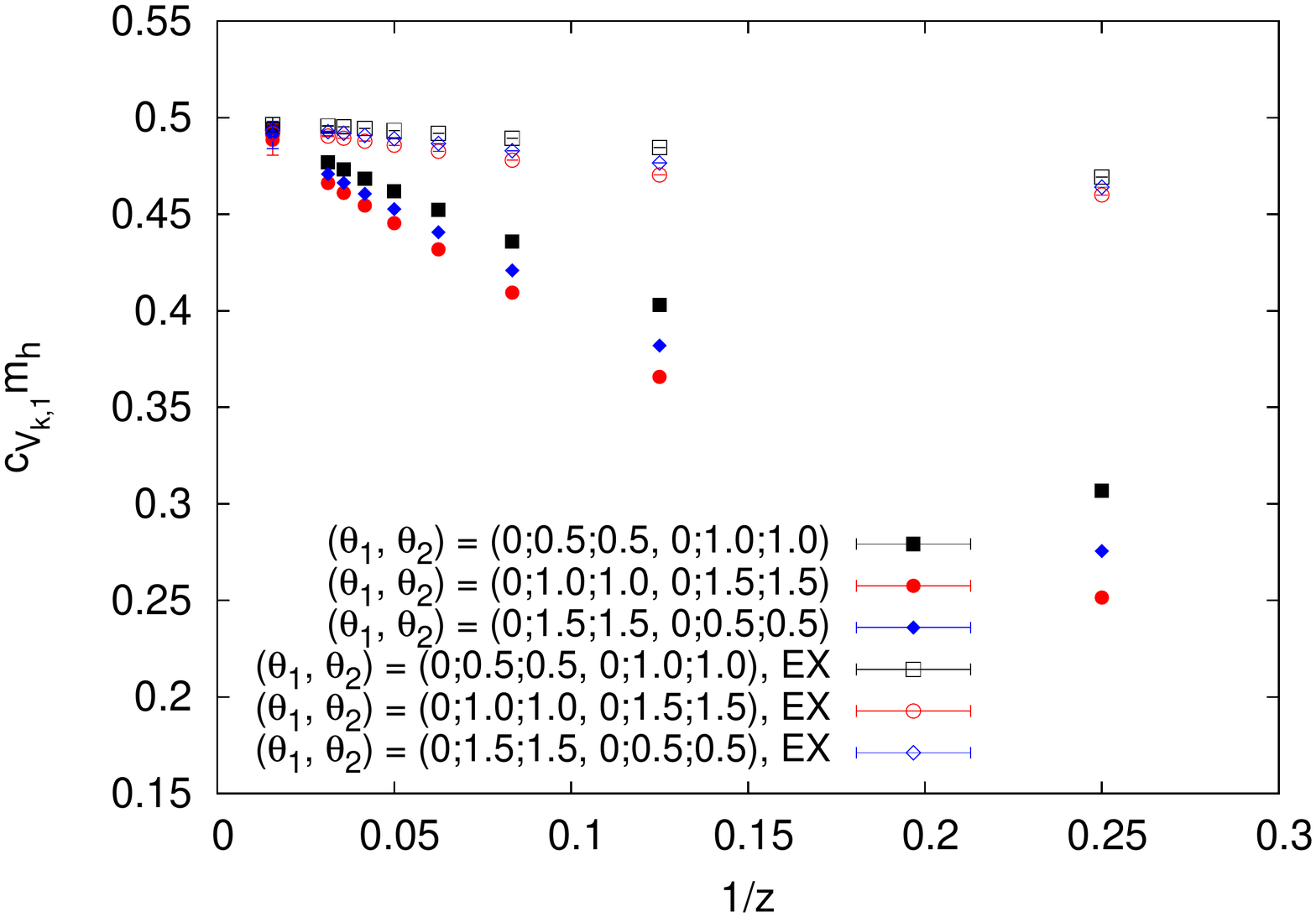} & 
\hspace{-1.8cm}\includegraphics[width=9.75cm]{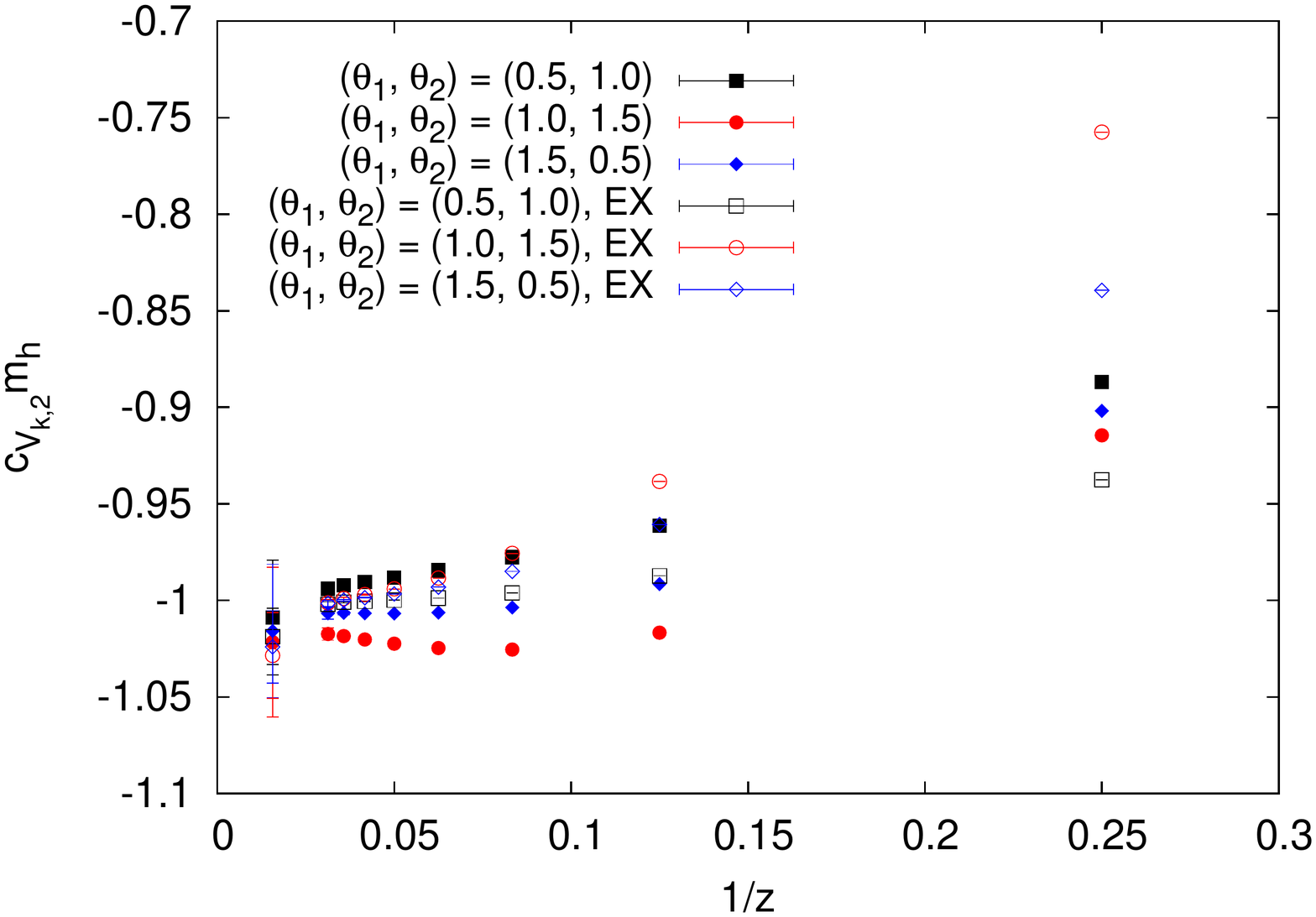} \\
\vspace{.4cm}
\includegraphics[width=9.75cm]{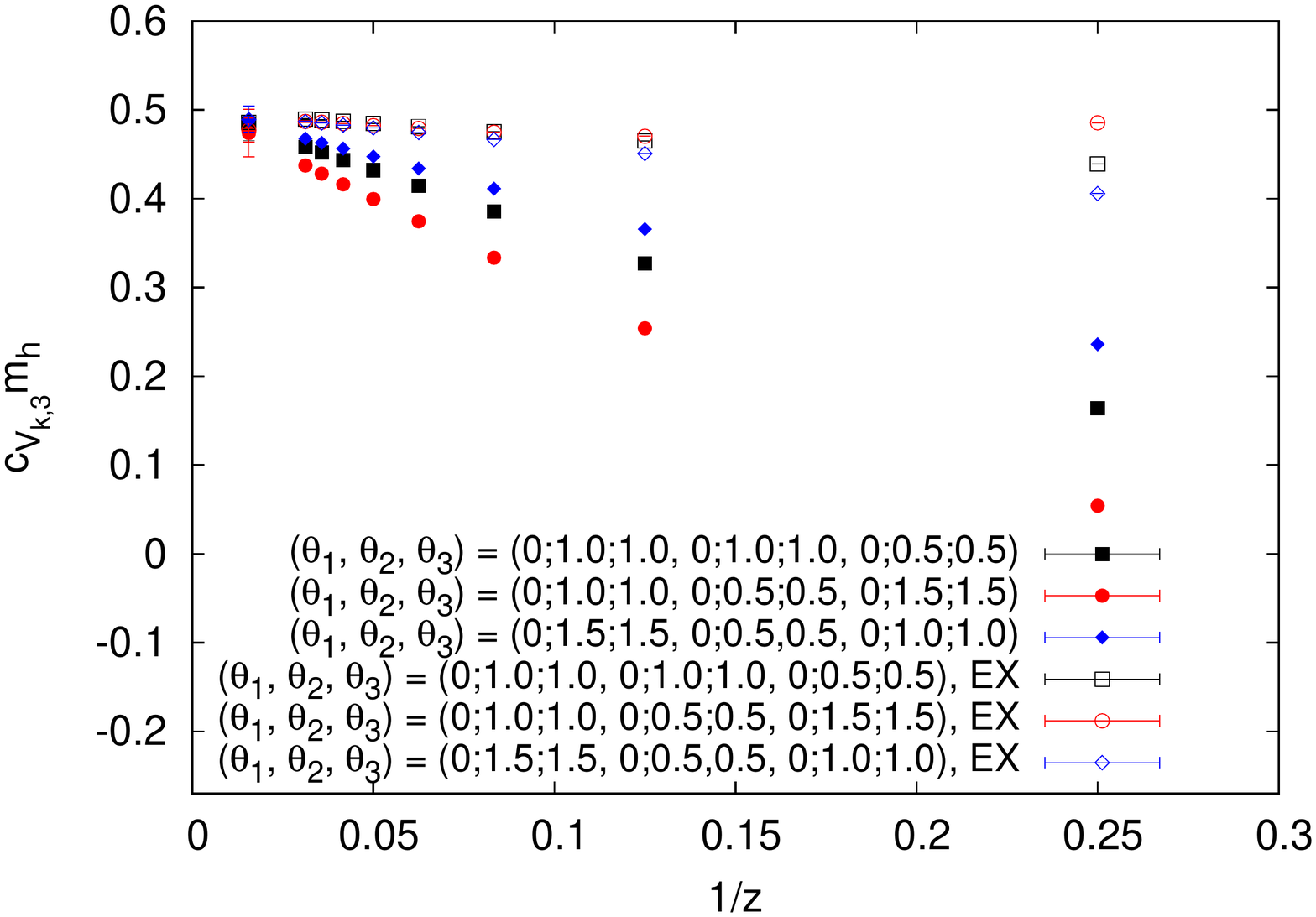} & 
\hspace{-1.8cm}\includegraphics[width=9.75cm]{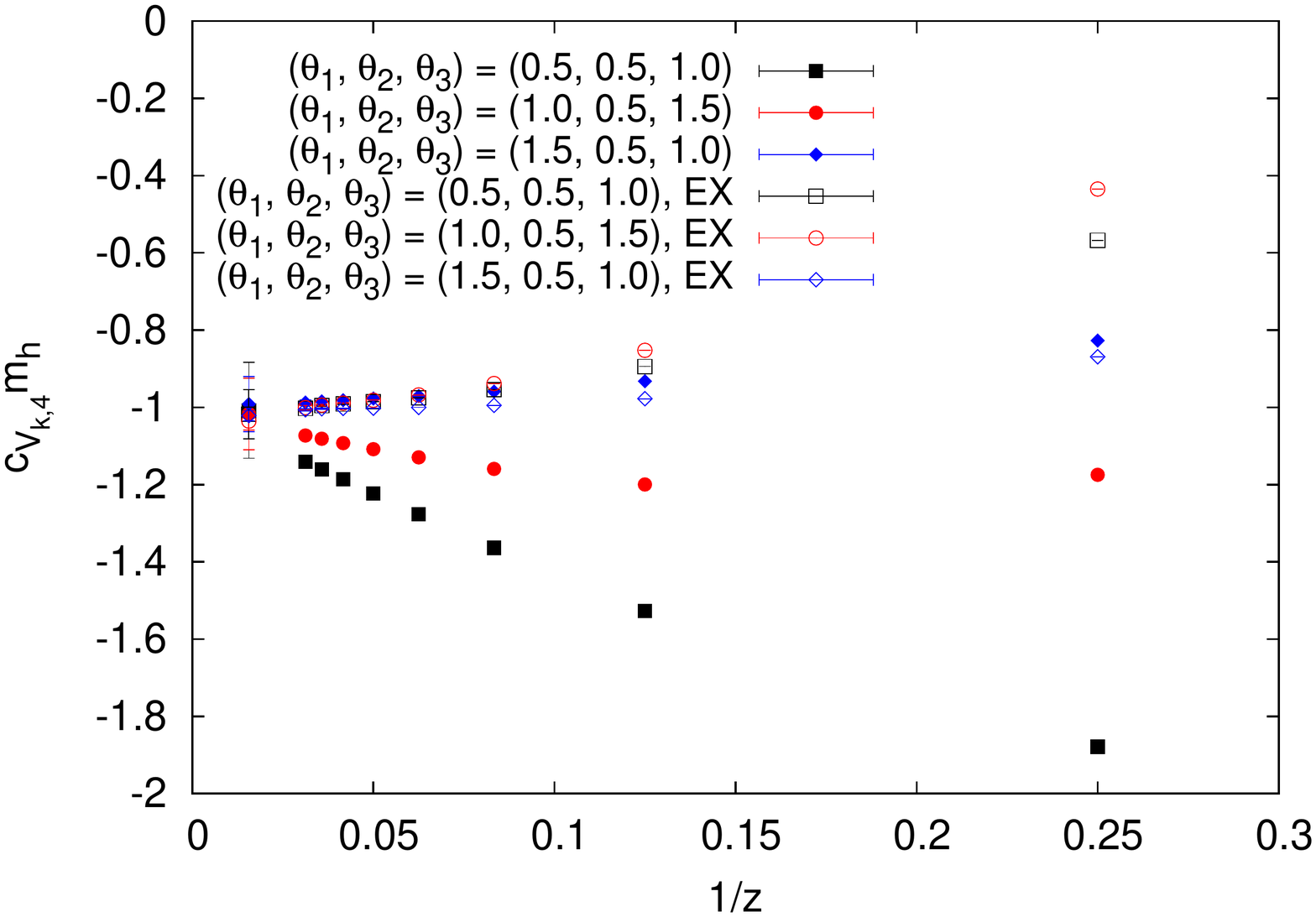} \\
\vspace{.4cm}
\includegraphics[width=9.75cm]{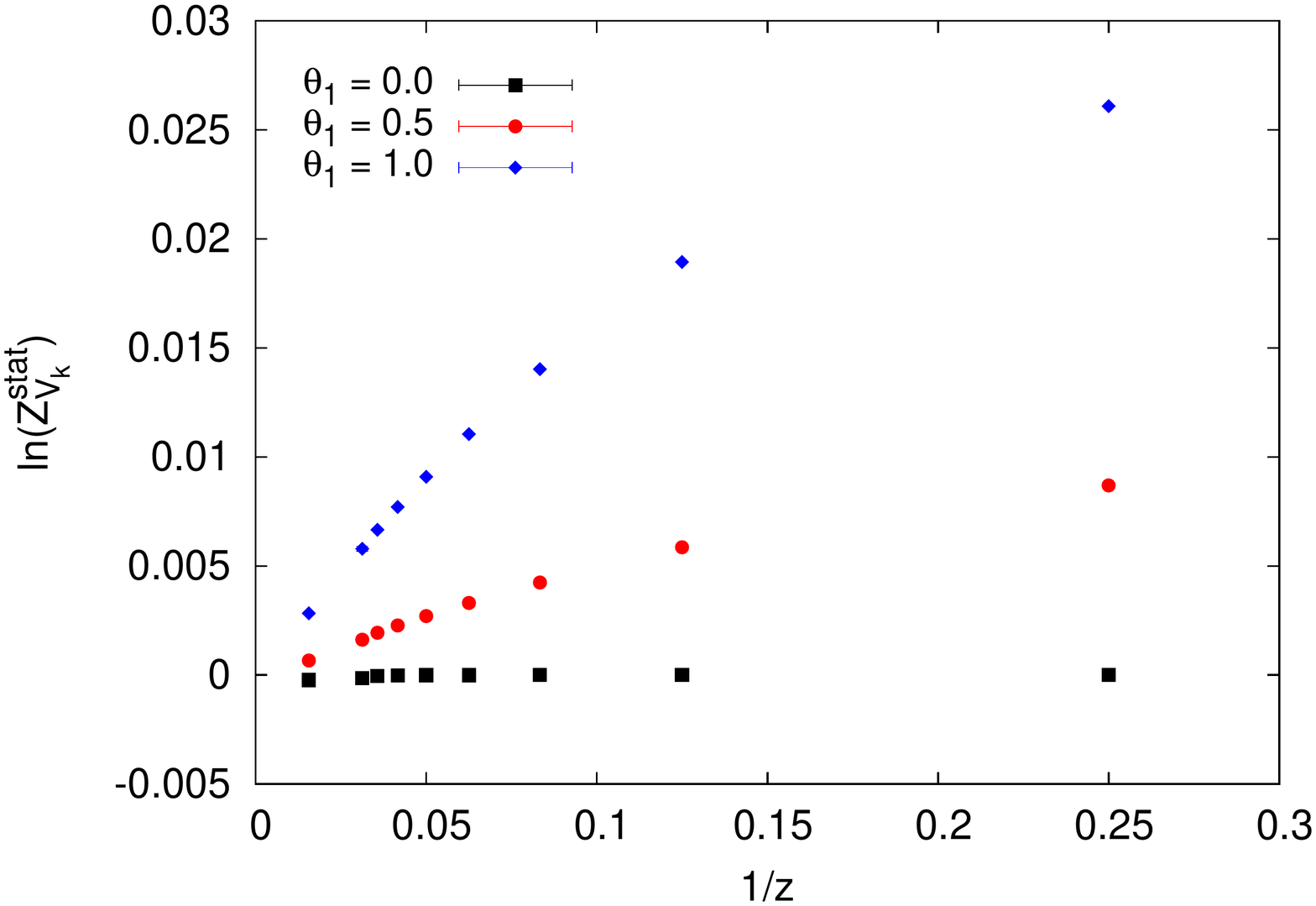} &
\hspace{-1.8cm}\includegraphics[width=9.75cm,height=6.68cm]{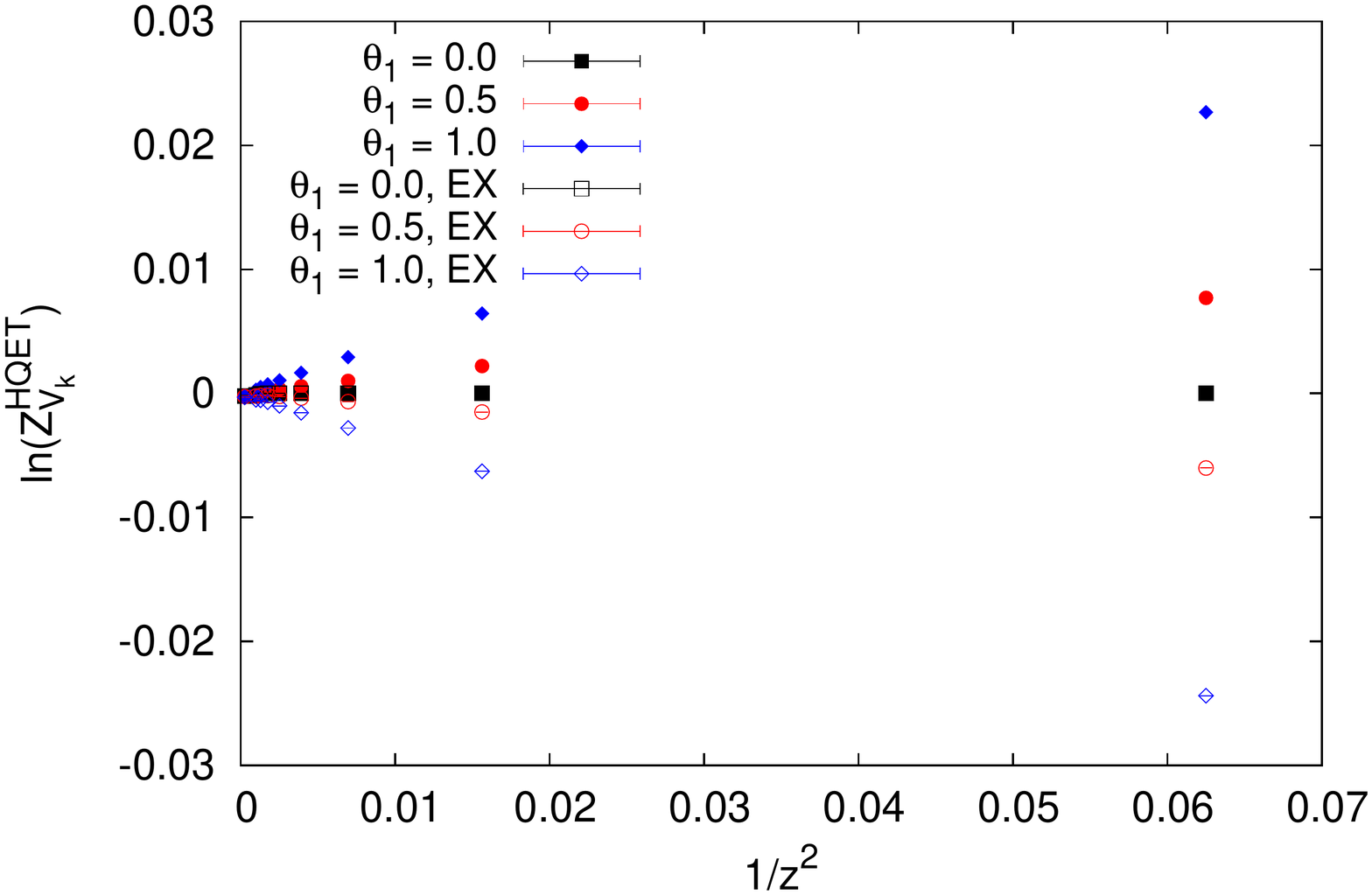}
\end{array}$
\caption{Tree-level continuum results for the parameters of the spatial component of the vector current. }
\label{fig:Vkplots}
\end{figure}
\begin{figure}[h!]
\hspace{-2.cm}
$\begin{array}{cc}
\includegraphics[width=9.75cm]{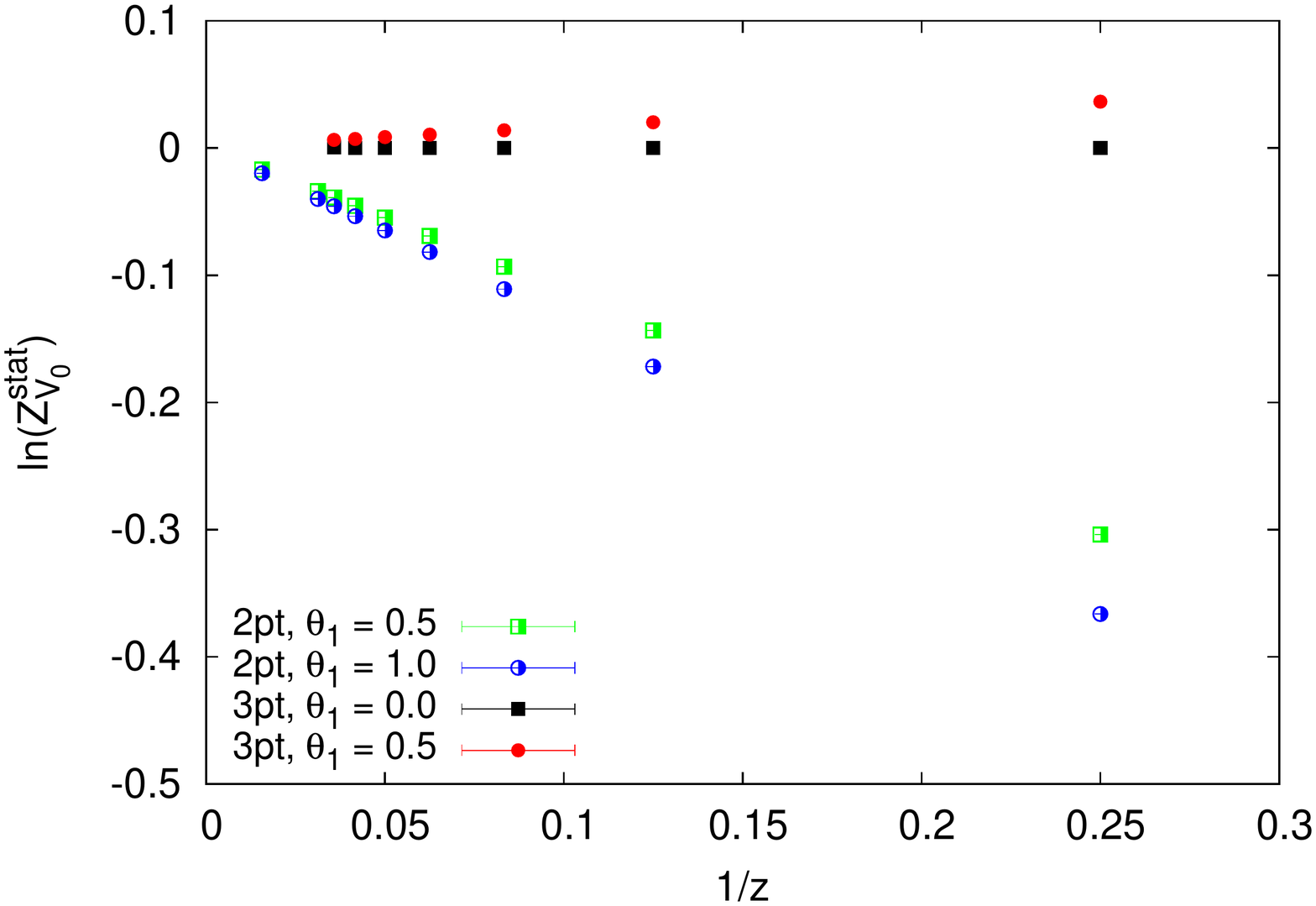} & 
\hspace{-1.8cm}\includegraphics[width=9.75cm,height=6.68cm]{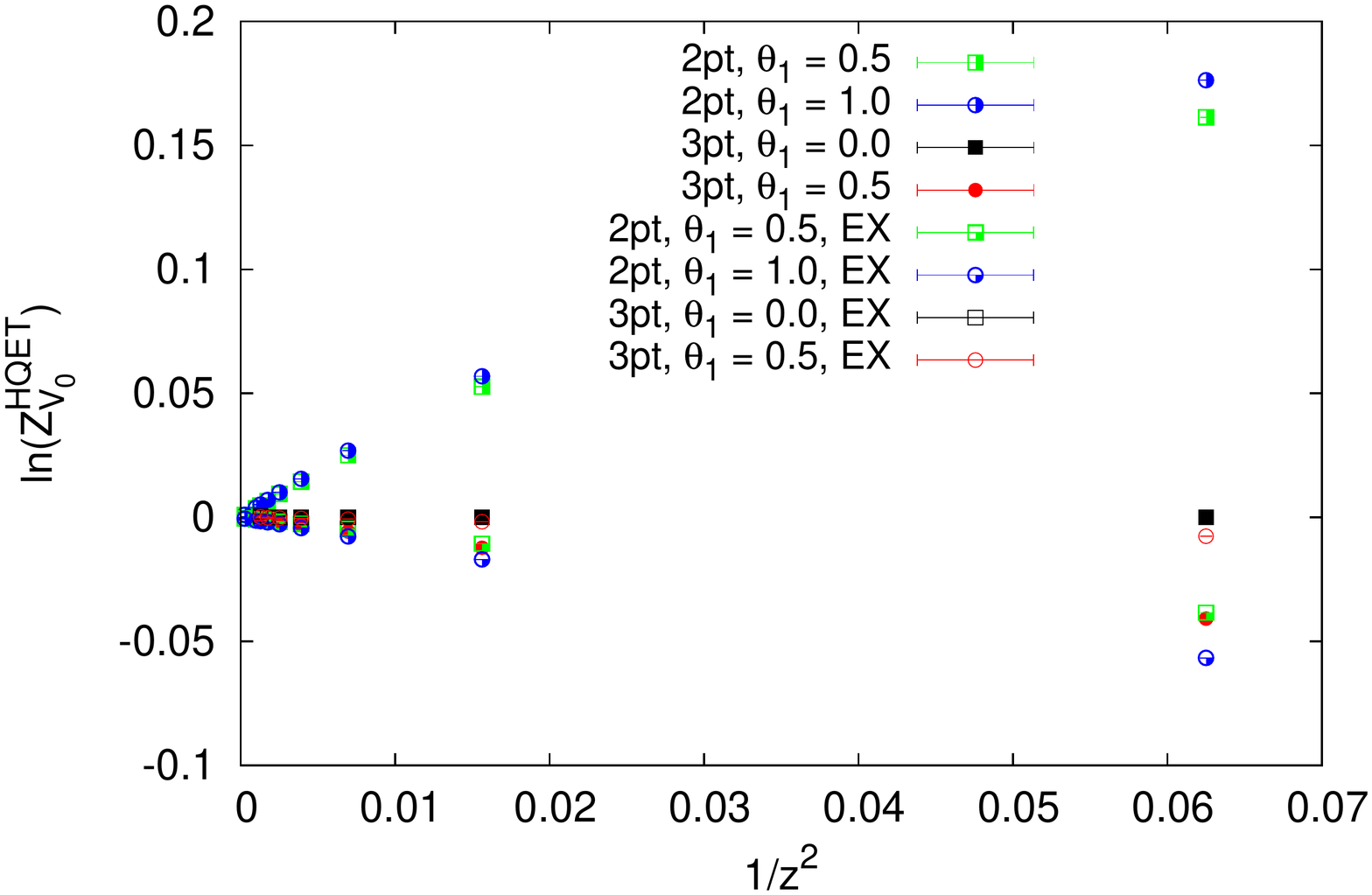}
\end{array}$
\caption{Comparison of tree-level continuum results for the renormalization constant of the 
temporal component of the vector current using either two-point correlation functions only (2pt), 
see eq.~\ref{e:Phi14old}, or also three-point functions (3pt), as in our standard definition 
of $\Phi_{14}^\qcd(\T_1)$.
}
\label{fig:V0-2and3}
\end{figure}

\section{Conclusions} 
\label{sec:concl}
We have presented a full set of matching conditions between HQET at 
$\Order{1/m_{\rm h}}$ and QCD for all the components of the axial and 
vector currents. This is the first time such an extended matching problem 
is formulated and solved on the lattice. 
The matrix of the linear system of equations to be solved in order to determine the 19 HQET parameters 
has a simple block structure. Therefore, once the 3 parameters of the 
Lagrangian have been computed, the system can then be solved independently 
for the parameters of each of the additional currents. 
The matching between QCD and HQET is performed with suitable observables 
in a finite volume with SF boundary conditions and exploits different kinematic 
situations by choosing appropriate twist angles $\T$.
We have evaluated and solved the resulting  matching equations at tree-level of perturbation theory and in the continuum limit. Thus, we have 
tested the feasibility of this strategy and demonstrated the 
sensitivity to all parameters (except for $\omega_\spin$, which does not 
enter at tree-level with vanishing background field).

In our tree-level computation we have also analysed the size 
of higher-order corrections (in $1/\mh$), which show up through the
dependence of the parameters on higher powers of $1/z$ and on the
 $\T$ angles. The block structure of the system of 
equations, which is further simplified at tree-level,
allowed us to independently investigate subsets of the 19 parameters.
In this way, we have identified a good kinematical 
setup, i.e., a choice of the $\T$ angles, to be adopted in the 
non-perturbative matching.
In view of a numerical 
implementation of that, the choice should also aim at 
minimizing the number of quark propagators to be computed or stored, i.e., 
at minimizing the number of combinations of $\T$ angles.

Remarkably, we find that in most of the cases the linear behaviour sets in already  
for $z\geq 8$, with slopes that are of $\Order{1}$. This is quite satisfactory, 
as it means that for the non-perturbative setup chosen by the 
ALPHA Collaboration~\cite{HQET:Nf2param1m}, with  $L \approx 0.5$ fm and $z$  
around 13 at the b-quark mass, higher-order corrections are suppressed by a 
factor of about 10.

Solving the matching problem is a necessary step towards a 
precise computation, within the framework of lattice HQET, of several 
phenomenologically relevant form factors describing semileptonic decays 
of ${\rm B}_{({\rm s})}$-mesons. The matching strategy presented here
is currently investigated in further detail at the one-loop order 
in perturbation theory~\cite{matchII} (see also ref.~\cite{beauty13:piotr} 
for partial results), and shall then be applied non-perturbatively in
numerical simulations in the near future and for the complete set of 19 matching
coefficients. We use Wilson fermions but, since the matching conditions
are espressed in terms of renormalized quantities, any regularization 
could in principle be used. Past experience shows that very fine lattice
spacings and good precision could be reached with Wilson 
fermions~\cite{HQET:param1m,HQET:Nf2param1m}.

\vskip1.5em
\noindent
{\bf Acknowledgements.}
We thank Rainer Sommer and Piotr Korcyl for many helpful discussions and comments, 
and for carefully reading the manuscript.
This work was partially supported by the Spanish Minister of 
Education and Science, project RyC-2011-08557 (M.~D.~M.),  by the grant HE~4517/3-1
of the Deutsche Forschungsgemeinschaft (J.~H.) and by the REA of the European Union
under Grant Agreement number PITN-GA-2009-238353, ITN STRONGnet (D.~H.).

\vspace{-0.1cm}
\begin{appendix}

\section{Explict form of the matching matrix for $\lag{HQET}$, $A_0$ and $A_k$}
\label{sec:a1}

In this appendix we give the explicit definition of the various $\deta{i}(\Tq)$ 
and $\dphi{i}{j}(\Th,\Tq)$ used in the HQET expansions in section~\ref{sec:s3}.

$\Phi^\hqet_1(\T_1)$: 
The quantities in the HQET expansion, eq.~\ref{e:mbare}, are 
\bea
    \deta{1}(\Tq)  & \equiv & 
    - L \tilde{\partial}_0 \left.\ln\left(-\fa^\stat(x_0,\Tq)\right)\right\vert_{x_0=T/2}\;, 
    \nonumber\\[2mm]
    \dphi{1}{1} & \equiv & L \;, \nonumber\\[2mm] 
    \dphi{1}{2}(\Tq,\Th) & \equiv &
    - L \tilde{\partial}_0 \left.\left( \frac{\fa^\kin(x_0,\Tq,\Th)}{\fa^\stat(x_0,\Tq)} \right)
    \right\vert_{x_0=T/2}\;, 
    \nonumber\\[2mm]
    \dphi{1}{3}(\Tq) & \equiv &
    - L \tilde{\partial}_0 \left.\left( \frac{\fa^\spin(x_0,\Tq)}{\fa^\stat(x_0,\Tq)} \right)
    \right\vert_{x_0=T/2} \;,
    \nonumber\\[2mm]
    \dphi{1}{4}(\Tq,\Th) & \equiv &
    - L \tilde{\partial}_0 \left.\left( \frac{f_{\rm A_{0,1}}(x_0,\Tq,\Th)}{\fa^\stat(x_0,\Tq)} \right)
    \right\vert_{x_0=T/2} \;.
\eea

$\Phi^\hqet_2\Atwo$: The way correlators enter in the HQET expansion, eq.~\ref{e:Phi2}, is
\bea
   \deta{2}(\Tq)    & \equiv & \ln F_1^\stat(\Tq)\;,  
   \nonumber \\[2mm]
   \dphi{2}{2}(\Tq,\Th) & \equiv & \frac{F_1^\kin(\Tq,\Th)}{F_1^\stat(\Tq)}\;.
\eea

$\Phi^\hqet_3(\T_1)$: The way correlators enter in the HQET expansion, eq.~\ref{e:Phi3}, is
\bea
   \dphi{3}{3}(\Tq) & \equiv & \frac{F_1^\spin(\Tq)}{F_1^\stat(\Tq)} \;.
\eea

$\Phi^\hqet_4\Atwo$: The way correlators  enter in the HQET expansion, eq.~\ref{e:Phi4}, is
\bea
   \deta{4}(\Tq)        & \equiv & \ln f_{\rm A_0}^\stat(T/2,\Tq)\;,
   \nonumber \\[2mm]
   \dphi{4}{2}(\Tq,\Th) & \equiv & \frac{f_{\rm A_0}^\kin(T/2,\Tq,\Th)}{f_{\rm A_0}^\stat(T/2,\Tq)}\;,
   \nonumber \\[2mm]
   \dphi{4}{3}(\Tq)     & \equiv & \frac{f_{\rm A_0}^\spin(T/2,\Tq)}{f_{\rm A_0}^\stat(T/2,\Tq)}\;,
   \nonumber \\[2mm]
   \dphi{4}{4}(\Tq,\Th) & \equiv & \frac{f_{\rm A_{0,1}}(T/2,\Tq,\Th)}{f_{\rm A_0}^\stat(T/2,\Tq)}\;.
\eea

$\Phi^\hqet_5\Athree$: The way correlators  enter in the HQET expansion, eq.~\ref{eq:phi5hqet}, is
\bea
   \dphi{5}{2}(\Tq,\Th) & \equiv & \dphi{4}{2}(\Tq,\Th)\;,
   \nonumber \\[2mm]
   \dphi{5}{4}(\Tq,\Th) & \equiv & \dphi{4}{4}(\Tq,\Th)\;,
   \nonumber \\[2mm]
   \dphi{5}{5}(\Tq,\Th) & \equiv & \frac{f_{\rm A_{0,2}}(T/2,\Tq,\Th)}{f_{\rm A_0}^\stat(T/2,\Tq)}\;.
\eea

$\Phi^\hqet_6(\T_1)$: The way correlators  enter in the HQET expansion, eq.~\ref{e:Phi6}, is
\bea
   \deta{6}(\Tq)   & \equiv & \ln\left(\frac{-\fa^\stat(T/2,\Tq)}{\sqrt{F_1^\stat(\Tq)}} \right)\;, 
   \nonumber \\[2mm]
   \dphi{6}{2}(\Tq,\Th) & \equiv & \frac{\fa^\kin(T/2,\Tq,\Th)}{\fa^\stat(T/2,\Tq)}  
                                - \frac{1}{2} \frac{F_1^\kin(\Tq,\Th)}{F_1^\stat(\Tq)}\;,
   \nonumber \\[2mm]
   \dphi{6}{3}(\Tq)     & \equiv & \frac{\fa^\spin(T/2,\Tq)}{\fa^\stat(T/2,\Tq)} 
                                - \frac{1}{2} \frac{F_1^\spin(\Tq)}{F_1^\stat(\Tq)}\;,
   \nonumber \\[2mm]
   \dphi{6}{4}(\Tq,\Th) & \equiv & \dphi{4}{4}(\Tq,\Th)\;.
\eea

After the temporal component, we now turn our attention to the spatial components of the axial 
current. 

For $\Phi^\hqet_7\Atwo$ in eq.~\ref{e:Phi7}:
\bea
    \deta{7}(\Tq)  & \equiv & 
    \ln f^\stat_{\vec{\rm A}}(T/2,\Tq)\;,
    \nonumber\\[2mm]
    \dphi{7}{2}(\Tq,\Th) & \equiv &
    \frac{f^\kin_{\vec{\rm A}}(T/2,\Tq,\Th)}{f^\stat_{\vec{\rm A}}(T/2,\Tq)}\;,
    \nonumber\\[2mm]
    \dphi{7}{3}(\Tq) & \equiv &
    \frac{f^\spin_{\vec{\rm A}}(T/2,\Tq)}{f^\stat_{\vec{\rm A}}(T/2,\Tq)}\;,
    \nonumber\\[2mm]
    \dphi{7}{7}(\Tq,\Th) & \equiv &
    \sum_{k}\frac{f_{{\rm A}_{k,1}}(T/2,\Tq,\Th)}{f^\stat_{\vec{\rm A}}(T/2,\Tq)}\;,
    \nonumber\\[2mm]
    \dphi{7}{8}(\Tq,\Th) & \equiv &
    \sum_k\frac{f_{{\rm A}_{k,2}}(T/2,\Tq,\Th)}{f^\stat_{\vec{\rm A}}(T/2,\Tq)}\;.
\eea

For $\Phi^\hqet_8\Atwo$ in eq.~\ref{e:Phi8}:
\bea
    \deta{8}(\Tq)  & \equiv & 
    \ln k^{1,\stat}_{\rm A_2}(T/2,\Tq)\;,
    \nonumber\\[2mm]
    \dphi{8}{2}(\Tq,\Th) & \equiv &
    \frac{k^{1,\kin}_{\rm A_2}(T/2,\Tq,\Th)}{k^{1,\stat}_{\rm A_2}(T/2,\Tq)}\;,
    \nonumber\\[2mm]
    \dphi{8}{3}(\Tq) & \equiv &
    \frac{k^{1,\spin}_{\rm A_2}(T/2,\Tq)}{k^{1,\stat}_{\rm A_2}(T/2,\Tq)}\;,
    \nonumber\\[2mm]
    \dphi{8}{7}(\Tq,\Th) & \equiv &
    \frac{k^1_{{\rm A}_{2,1}}(T/2,\Tq,\Th)}{k^{1,\stat}_{\rm A_2}(T/2,\Tq)}\;,
    \nonumber\\[2mm]
    \dphi{8}{8}(\Tq,\Th) & \equiv &
    \frac{k^1_{{\rm A}_{2,2}}(T/2,\Tq,\Th)}{k^{1,\stat}_{\rm A_2}(T/2,\Tq)}\;.
\eea
Note that the contribution from $\dphi{8}{8}(\Tq,\Th)$ vanishes at tree-level,
because of its Dirac structure, for any choice of $\Th$, $\Tq$.

For $\Phi^\hqet_9\Athree$ in eq.~\ref{e:Phi9}:
\bea
    \dphi{9}{2}(\Tq,\Th) & \equiv & \dphi{7}{2}(\Tq,\Th)\;, 
    \nonumber\\[2mm]
    \dphi{9}{7}(\Tq,\Th) & \equiv & \dphi{7}{7}(\Tq,\Th) \;,
    \nonumber\\[2mm]
    \dphi{9}{8}(\Tq,\Th) & \equiv & \dphi{7}{8}(\Tq,\Th) \;,
    \nonumber\\[2mm]
    \dphi{9}{9}(\Tq,\Th) & \equiv & 
    \sum_k \frac{f_{{\rm A}_{k,3}}(T/2,\Tq,\Th)}{f^\stat_{\vec{\rm A}}(T/2,\Tq)}\;,
    \nonumber\\[2mm]
    \dphi{9}{10}(\Tq,\Th) & \equiv & 
    \sum_k \frac{f_{{\rm A}_{k,4}}(T/2,\Tq,\Th)}{f^\stat_{\vec{\rm A}}(T/2,\Tq)}\;.
\eea

For $\Phi^\hqet_{10}\Athree$ in eq.~\ref{e:Phi10}:
\bea
    \dphi{10}{2}(\Tq,\Th) & \equiv & \dphi{8}{2}(\Tq,\Th)\;,
    \nonumber\\[2mm]
    \dphi{10}{7}(\Tq,\Th) & \equiv & \dphi{8}{7}(\Tq,\Th)\;,
    \nonumber\\[2mm]
    \dphi{10}{8}(\Tq,\Th) & \equiv & \dphi{8}{8}(\Tq,\Th)\;,
    \nonumber\\[2mm]
    \dphi{10}{9}(\Tq,\Th) & \equiv &
    \frac{k^1_{{\rm A}_{2,3}}(T/2,\Tq,\Th)}{k^{1,\stat}_{\rm A_2}(T/2,\Tq)}\;,
    \nonumber\\[2mm]
    \dphi{10}{10}(\Tq,\Th) & \equiv &
    \frac{k^1_{{\rm A}_{2,4}}(T/2,\Tq,\Th)}{k^{1,\stat}_{\rm A_2}(T/2,\Tq)}\;.
\eea
Again, the contribution from $\dphi{10}{8}(\Tq,\Th)$ and in addition from $\dphi{10}{10}(\Tq,\Th)$
vanish at tree-level.

For $\Phi^\hqet_{11}(\T_1)$ in eq.~\ref{e:Phi11}:
\bea
   \deta{11}(\Tq)          & \equiv & \ln \left( {{J_{{\rm A}_1}^{1\; \rm stat}(T/2,\T_{\ell},\T_{\ell'})}\over{\sqrt{K_1^{\ell\ell}(\T_\ell,\T_{\ell'})
                           \times F_1^\stat(\T_\ell)}}} \right)\;,
   \nonumber \\[2mm]
   \dphi{11}{2}(\Tq,\Th)   & \equiv &   \left(  {{J_{{\rm A}_1}^{1\; \rm kin}(T/2,\T_\ell,\T_{\ell'},\T_{\rm h})}\over
{{J_{{\rm A}_1}^{1\; \rm stat}(T/2,\T_\ell,\T_{\ell'})}}}-{{1}\over{2}} {{F_{1}^\kin(\T_\ell,\T_{\rm h})} \over {F_1^\stat(\T_\ell)}}\right)\;, 
   \nonumber \\[2mm]
   \dphi{11}{3}(\Tq)       & \equiv &   \left(  {{J_{{\rm A}_1}^{1\; \rm spin}(T/2,\T_\ell,\T_{\ell'})}\over
{{J_{{\rm A}_1}^{1\; \rm stat}(T/2,\T_\ell,\T_{\ell'})}}}-{{1}\over{2}} {{F_1^\spin(\T_\ell)} \over {F_1^\stat(\T_\ell)}}\right)\;,
   \nonumber \\[2mm]
   \dphi{11}{7}(\Tq,\Th)   & \equiv &  {{J_{{{\rm A_1}},1}^1(T/2,\T_\ell,\T_{\ell'},\T_{\rm h})}\over{J_{{\rm A}_1}^{1\; \rm stat}(T/2,\T_\ell,\T_{\ell'})}}\;,
   \nonumber \\[2mm]
   \dphi{11}{8}(\Tq,\Th)   & \equiv & {{J_{{{\rm A_1}},2}^1(T/2,\T_\ell,\T_{\ell'},\T_{\rm h})}\over{J_{{\rm A}_1}^{1\; \rm stat}(T/2,\T_\ell,\T_{\ell'})}}\;.
\eea

The terms in the alternative observable $\Phi^{\prime\,\hqet}_{11}(\T_1)$ in eq.~\ref{e:Phi112pt} explicitly read
\bea
   \deta{11}^\prime(\Tq)          & \equiv & \ln\left(\frac{\fak^\stat(T/2,\Tq)}{\sqrt{F_1^\stat(\Tq)}} \right)\;, 
   \nonumber \\[2mm]
   \dphi{11}{\prime\,2}(\Tq,\Th)   & \equiv & \frac{\fak^\kin(T/2,\Tq,\Th)}{\fak^\stat(T/2,\Tq)} 
                                    - \frac{1}{2} \frac{F_1^\kin(\Tq,\Th)}{F_1^\stat(\Tq)}\;,
   \nonumber \\[2mm]
   \dphi{11}{\prime\,3}(\Tq)       & \equiv & \frac{\fak^\spin(T/2,\Tq)}{\fak^\stat(T/2,\Tq)} 
                                    - \frac{1}{2} \frac{F_1^\spin(\Tq)}{F_1^\stat(\Tq)}\;,
   \nonumber \\[2mm]
   \dphi{11}{\prime\,7}(\Tq,\Th)   & \equiv & \dphi{7}{7}(\Tq,\Th)\;,
   \nonumber \\[2mm]
   \dphi{11}{\prime\,8}(\Tq,\Th)   & \equiv & \dphi{7}{8}(\Tq,\Th)\;.
\eea

\section{Correlation functions and observables for the matching of the vector current}
\label{sec:a2}

We introduce the following boundary-to-boundary, boundary-to-bulk and three-point correlation functions
\bea
   F^{\light\light}_1(\vec{\theta}_{\ell},\vec{\theta}_{\ell'}) & = &
     -{a^{12} \over 2L^6}\sum_{\vecu,\vecv,\vecy,\vecz}
     \left\langle
     \zetabarprime_\light(\vecu)\gamma_5\zetaprime_{\ell'}(\vecv)\,
     \zetabar_{\ell'}(\vecy)\gamma_5\zeta_\light(\vecz)
     \right\rangle\,, 
\\
\kv0
(x_0,\theta_\ell,\thh) & = & i{a^6 \over 6}\sum_{k}\sum_{\vecy,\vecz}\,
  \left\langle
  (\vimpr)_0(x)\,\zetabar_{\rm b}(\vecy)\gamma_k\zeta_\light(\vecz)
  \right\rangle  \,, \label{e_kv0}
\\
\kvk
(x_0,\theta_\ell,\thh) & = & -{a^6 \over 6}\sum_{k}\sum_{\vecy,\vecz}\,
  \left\langle
  (\vimpr)_k(x)\,\zetabar_{\rm b}(\vecy)\gamma_k\zeta_\light(\vecz)
  \right\rangle  \,, \label{e_kv}
\\
\kvuu
(x_0,\theta_\ell,\thh) & = & -{a^6 \over 2}\sum_{\vecy,\vecz}\,
  \left\langle
  (\vimpr)_1(x)\,\zetabar_{\rm b}(\vecy)\gamma_1\zeta_\light(\vecz)
  \right\rangle  \,, \label{e_kv11}
\\
\f1v0
(x_0,\T_{\ell},\T_{\ell'},\thh) & = & -{{a^{15}}\over{2L^6}}\sum_{\bf u,v,y,z,x} 
  \langle \zetabar'_{\ell'}({\bf u}) \gamma_5 \zeta'_{\ell} ({\bf v})
  (V_{\rm I})_0(x) \zetabar_{\rm b} ({\bf z})\gamma_5 \zeta_{\ell'} ({\bf y})\rangle \;,
\label{e_f1v0}
\eea
where at least one component of the $\T$ angles in $\kv0$ should not be zero.
With these correlators at hand we build the following matching observables
\def\phiseparator{8mm}
\bd
  \begin{array}{lcll}
  \Phi_{12}^\qcd\Atwo & \equiv & \displaystyle
      \ln\left(\frac{k_{\rm V_0}(x_0,\T_1,\T_1)}{k_{\rm V_0}(x_0,\T_2,\T_2)}\right)\,, 
  & (T=L,\ x_0=T/2)\,,
  \\[\phiseparator]
  \Phi_{13}^\qcd\Athree & \equiv & \displaystyle
      \ln\left(\frac{k_{\rm V_0}(x_0,\T_1,\T_2)}{k_{\rm V_0}(x_0,\T_1,\T_3)}\right) \,,
  & (T=L,\ x_0=T/2)\,,
  \\[\phiseparator]
  \Phi_{14}^\qcd(\T_1) & \equiv & \displaystyle 
      \ln\left(\frac{F_{\rm V_0}(x_0,\T_1,\T_1,\T_1)}{\sqrt{F_1(\T_1,\T_1) \times F^{\light\light}_1(\T_1,\T_1)}} \right)\,,
  & (T=L,\ x_0=T/2)\,,
  \\[\phiseparator]
  \Phi_{15}^\qcd\Atwo & \equiv & \displaystyle 
      \ln\left(\frac{k_{\vec{\rm V}}(x_0,\T_1,\T_1)}{k_{\vec{\rm V}}(x_0,\T_2,\T_2)}\right)\,,
  & (T=L,\ x_0=T/2)\,,
  \\[\phiseparator]
  \Phi_{16}^\qcd\Atwo & \equiv & \displaystyle
      \ln\left(\frac{k^1_{\rm V_1}(x_0,\T_1,\T_1)}{k^1_{\rm V_1}(x_0,\T_2,\T_2)}\right)\,,
  & (T=L,\ x_0=T/2)\,,
  \\[\phiseparator]
  \Phi_{17}^\qcd\Athree & \equiv & \displaystyle 
      \ln\left(\frac{k_{\vec{\rm V}}(x_0,\T_1,\T_2)}{k_{\vec{\rm V}}(x_0,\T_1,\T_3)}\right)\,,
  & (T=L,\ x_0=T/2)\,,
  \\[\phiseparator]
  \Phi_{18}^\qcd\Athree & \equiv & \displaystyle 
      \ln\left(\frac{k^1_{\rm V_1}(x_0,\T_1,\T_2)}{k^1_{\rm V_1}(x_0,\T_1,\T_3)}\right)\,,
  & (T=L,\ x_0=T/2)\,,
\end{array}
\ed
  \\[\phiseparator]
\bd
 \begin{array}{lcll}
\Phi_{19}^\qcd(\T_1) &  \equiv  &\displaystyle
      \ln\left(\frac{-k_{\vec{\rm V}}(x_0,\T_1,\T_1)}{\sqrt{K_1(\T_1,\T_1)}}\right)\,, 
  & (T=L,\ x_0=T/2)\;.
\end{array}
\ed
An alternative for $\Phi_{14}^{\rm QCD}(\T_1)$, employing only two-point 
functions, is provided by
\be
\Phi_{14}^{\prime\,\qcd}(\T_1)  \equiv  \displaystyle
\ln\left(\frac{k_{\rm V_0}(x_0,\T_1,\T_1)}{\sqrt{K_1(\T_1,\T_1)}}\right)\,, 
\qquad  (T=L,\ x_0=T/2)\;.
\label{e:Phi14old}
\ee
In choosing the $\T$ angles for the results presented in this paper, 
we have made use of the fact that the correlators
$k_{\rm V_{1,2}}^1$ and $k_{\rm V_{1,4}}^1$ vanish at tree-level, if the
$x$-component of all the $\T$'s are set to zero.
This option helps in  simplifying the system of equations and in making it 
entirely solvable by backward substitution.

The HQET expansion of the observables above can very easily be derived
from the expansion of the quantities $\Phi_4^{\qcd}$, \ldots, $\Phi_{11}^{\qcd}$ in 
appendix~\ref{sec:a1}
through the obvious replacements of correlators.

\end{appendix}

\bibliographystyle{JHEP}       
\bibliography{mainbib_alpha}   

\end{document}